\documentclass[longauth]{aa}  
\usepackage{graphicx}
\usepackage{txfonts}
\usepackage[colorlinks=true,allcolors=blue]{hyperref}
\usepackage{newtxtext,newtxmath}
\usepackage[T1]{fontenc}
\usepackage{amsmath}
\usepackage{amssymb}
\usepackage{color}
\usepackage{hyperref}
\usepackage{pdflscape}
\newcommand{\orcid}[1]{\href{https://orcid.org/#1}{\includegraphics[width=10pt]{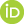}}}
\usepackage{lipsum} 
\usepackage{leftidx}
\usepackage{float}
\usepackage{tabularx}
\usepackage[font=footnotesize]{caption}
\usepackage{multirow}
\definecolor{forestgreen}{rgb}{0.0, 0.7, 0.1}

\usepackage[switch]{lineno}

\begin{document} 
\title{ZTF SN Ia DR2 follow-up: Characterization of subluminous \\
Type Ia supernovae in the ZTF DR2 full sample}
\titlerunning{ZTF DR2 subluminous SNe Ia}
\authorrunning{Alburai et al.}

\author{Alaa Alburai\inst{1,2}\thanks{alburai@ice.csic.es}\orcid{0009-0007-2731-5562},
Llu\'is Galbany\inst{1,2}\orcid{0000-0002-1296-6887},
Umut Burgaz\inst{3}\orcid{0000-0003-0126-3999},
Georgios Dimitriadis\inst{4}\orcid{0000-0001-9494-179X},
Joel Johansson\inst{5}\orcid{0000-0001-5975-290X},\\
Mat Smith\inst{4}\orcid{},
Ramon Sanfeliu\inst{1,2}\orcid{},
Sandra Guerra\inst{1,2}\orcid{},
Tom\'as M\"uller-Bravo\inst{3,6}\orcid{0000-0003-3939-7167},
Ariel Goobar\inst{5}\orcid{0000-0002-4163-4996},\\
Suhail Dhawan\inst{7}\orcid{0000-0002-2376-6979},
Young-Lo Kim\inst{8}\orcid{0000-0002-1031-0796},
Jakob Nordin\inst{9}\orcid{0000-0001-8342-6274},
Alice Townsend\inst{9}\orcid{0000-0001-6343-3362},
Jesper Sollerman\inst{11}\orcid{0000-0003-1546-6615},\\
Madeleine Ginolin\inst{10}\orcid{0009-0004-5311-9301},
Mickael Rigault\inst{10}\orcid{0000-0002-8121-2560},
Jacco H. Terwel\inst{3}\orcid{0000-0001-9834-3439},
Roger Smith\inst{12}\orcid{0000-0001-7062-9726},\\
Avery Wold\inst{13}\orcid{0000-0002-9998-6732},
Tracy X. Chen\inst{13}\orcid{0000-0001-9152-6224},
Theophile Jegou du Laz\inst{14}\orcid{0009-0003-6181-4526},
}

\institute{
Institute of Space Sciences (ICE, CSIC), Campus UAB, Carrer de Can Magrans, s/n, E-08193 Barcelona, Spain. \and
Institut d’Estudis Espacials de Catalunya (IEEC), E-08034 Barcelona, Spain. \and
School of Physics, Trinity College Dublin, The University of Dublin, Dublin 2, Ireland. \and
Department of Physics, Lancaster University, Lancaster, LA1 4YB, UK. \and
Oskar Klein Centre, Department of Physics, Stockholm University, SE-10691 Stockholm, Sweden. \and
Instituto de Ciencias Exactas y Naturales (ICEN), Universidad Arturo Prat, Chile. \and
School of Physics and Astronomy, University of Birmingham, Birmingham, UK. \and
Department of Astronomy \& Center for Galaxy Evolution Research, Yonsei University, Seoul 03722, Republic of Korea. \and
Institut f\"ur Physik, Humboldt-Universit\"at zu Berlin, Newtonstr. 15, 12489 Berlin, Germany.\and
Univ. Lyon, Univ. Claude Bernard Lyon 1, CNRS, IP2I Lyon/IN2P3, UMR 5822, F-69622 Villeurbanne, France. \and 
Oskar Klein Centre, Department of Astronomy, Stockholm University, SE-10691 Stockholm, Sweden. \and 
Caltech Optical Observatories, California Institute of Technology, Pasadena, CA  91125, USA. \and
IPAC, California Institute of Technology, 1200 E. California
Blvd, Pasadena, CA 91125, USA. \and
Division of Physics, Mathematics and Astronomy, California Institute of Technology, Pasadena, CA 91125, USA.}
\date{Received \today; accepted XXX}

\abstract
{
The Zwicky Transient Facility Data Release 2 (ZTF DR2) includes a total of 3,628 Type Ia supernovae (SNe~Ia), providing the largest and most complete sample of spectroscopically confirmed SNe~Ia at low redshift to date. In this paper, we present a photometric and spectroscopic analysis of 124 subluminous SNe~Ia, the largest sample of spectroscopically classified subluminous type Ia supernova observed with a single instrument, comprising 87 91bg-like, 12 86G-like, 18 04gs-like, and 7 02es-like events.
We complement the published DR2 \textsc{SALT2} light-curve parameters with new parameters obtained using template-based fits from \textsc{SNooPy}. Expansion velocities and pseudo-equivalent widths ($pEW$) of key spectral features are measured using \textsc{Spextractor}, and spectral averages are constructed for each subluminous subtype, binned by phase. We also analyze the host galaxy environments, both global and local, in terms of $g - z$ color, stellar mass, and directional light radius ($d_{DLR}$).
We find that all subluminous SNe~Ia (except the 02es-like subtype) are intrinsically red. This is evident by separating extrinsic from intrinsic color components. Since \textsc{SALT2} is not trained on subluminous SNe~Ia, it compensates for their redder colors by inflating the $c$ parameter, thus extending the luminosity-width relation to negative values of $x1$.
As expected, all subluminous SNe~Ia fall within the \textit{Cool} region of the Branch et al. (2006) diagram, with the exception of 02es-like events, which show lower \ion{Si}{ii}~$\lambda$5972 $pEW$ values. All subluminous subtypes tend to occur in more massive, redder host galaxies, and in the reddest local environments within their stellar mass bins. Notably, 91bg- and 86G-like SNe~Ia explode at significantly larger normalized galactocentric distances.
Finally, we identify the $pEW$ of the blended \ion{Ti}{ii}+\ion{Si}{ii}+\ion{Mg}{ii} absorption feature at 4300~\AA, along with $s_{BV}$, as robust and sufficient indicators for subclassifying subluminous SNe~Ia.}
\maketitle


\section{Introduction}

Type Ia supernovae (SNe Ia) are the explosive endings of massive stars and white dwarfs in binary stellar systems. As sources of new elements, they enrich the interstellar medium with freshly synthesized metals and release immense kinetic energy that drives galaxy evolution. SNe Ia  are some of the most luminous transient phenomena in the universe, typically reaching peak absolute magnitudes of about $-$19.5 mag. They are also relatively homogeneous showing variations in peak brightness of only 1$-$2 magnitudes (e.g. \citealt{1996AJ....112.2391H}).
Through empirical calibrations like the relations between peak brightness and post-maximum brightness decline (e.g., $\Delta m_{15}(B)$, the parameter describing magnitude fade after 15 days past maximum in $B$-band; \citealt{1993ApJ...413L.105P}), and peak brightness and color at maximum \citep{1996ApJ...473...88R, 1998A&A...331..815T}, this scatter is reduced to $\sim$ 0.1 mag. Due to their standardizable nature and the ability to detect them in ground-based observations even at high redshifts ($z$ $\sim$ 1.0), SNe Ia have been widely used in recent decades to refine measurements of cosmological parameters \citep{2014A&A...568A..22B,2024ApJ...973L..14D} and calculate the Hubble constant \citep{2022ApJ...934L...7R,2023A&A...679A..95G,2024ApJ...970...72U}. Notably, they were the key in the discovery of the accelerated expansion of the Universe, a breakthrough achieved by \cite{1998AJ....116.1009R} and \cite{1999ApJ...517..565P}. Despite being regarded as the gold standard for cosmological measurements, our understanding of the progenitor systems, explosion mechanisms, and the physical basis for the empirical relationships that make SNe Ia precise distance indicators remains under debate. Aside from a deeper understanding of the explosion model, additional advances demand further constraints on what type of progenitors lead to SNe Ia and also require improved control over reddening and a full quantification of peculiar (non-standardizable) subtypes.

The consensus is that a SN Ia arises from a carbon-oxygen (C-O) white dwarf (WD), which is supported by electron degeneracy pressure, experiencing a thermonuclear runaway \citep{1960ApJ...132..565H,2023RAA....23h2001L}. It is also generally accepted that SNe Ia originate from close binary systems. However, the specific characteristics of these systems and the explosion process are still debated \citep{2016IJMPD..2530024M,1991A&A...245..114K}. SNe Ia may originate from two main types of progenitor systems, which could be either two white dwarfs, known as a double degenerate (DD) system, or a single white dwarf with a non-degenerate companion, known as a single degenerate (SD) system. The most widely considered progenitor scenarios are the following: (i) Dynamical mergers of two carbon-oxygen (C-O) white dwarfs in a binary system, where the heat generated by the merging process ignites the thermonuclear explosion \citep{2017hsn..book.1237G}. (ii) Chandrasekhar-mass ($M_{Ch}$) explosions of a C-O white dwarf near the Chandrasekhar mass, with compressional heating at its center triggering ignition. This process occurs due to accretion from a donor star (eg. red giant star) with a mass below 7–8 solar masses \citep{2022ApJ...941L..33A,2010A&A...516A..47M}, a helium star, or even a tidally disrupted white dwarf from a DD system \citep{2004MNRAS.353..243P}. (iii) Sub-Chandrasekhar-mass detonations of a C-O white dwarf caused by the detonation of a thin helium layer on its surface, a scenario often referred to as "double detonations" \citep{2010ApJ...719.1067K,2014ApJ...797...46S}. (iv) Direct collisions of two white dwarfs in tertiary systems, where a third star induces eccentricity oscillations, enhancing gravitational-wave losses and finally causing the collision \citep{2018MNRAS.476.2905M}.

The debate about the four scenarios discussed above is ongoing, however, the connection between these progenitor scenarios and specific SN Ia subtypes is still unclear. For instance, extremely bright or faint SNe Ia may require progenitors or explosion mechanisms that differ from these scenarios. Investigating and establishing which progenitor configuration and explosion mechanism lead to SNe Ia with particular properties is crucial for advancing our understanding of stellar evolution and supernova cosmology. The presence of these peculiar over- and subluminous SNe Ia, and other heterogeneous objects that do not follow the empirical relations \citep{2017hsn..book..317T,2025A&A...694A..10D}, may challenge the picture of all SNe Ia coming from the same family of progenitors.

Unlike the over-luminous 91T-like SNe Ia, which primarily occur in actively star-forming late-type galaxies, the sub-luminous 91bg-like SNe Ia are predominantly observed in large elliptical or S0 galaxies with low star-formation rates, typically only a few times $10^{-9} M_{\odot}  \text{yr}^{-1}$ \citep{2002ApJ...568..779H,2004ApJ...613.1120G}. In particular, these SNe seem to break certain relations, such as those involving color characteristics and evolution, and exhibit peculiarities in their light curves. One of their main distinguishing features is the absence of a secondary maximum in the optical-red and NIR light curves, a characteristic commonly seen in normal SNe Ia \citep{2012PASA...29..434P}. They also show deviations from the luminosity-width relation \citep{1993ApJ...413L.105P}, which affects their reliability as cosmic distance indicators. Photometrically, these objects are fainter at peak brightness and have faster rise and decline rates in their light curves, placing them at the faint and fast-extreme end of the luminosity-width relation ($\Delta m_{15}(B)>1.7$ mag). The absence of either a shoulder or a secondary maximum in the NIR light curves can be explained by the earlier recombination of Fe III to Fe II. The Fe II line blanketing absorbs flux in the blue, which is then re-emitted at longer wavelengths, causing the first and second NIR peaks to merge into a single, slightly delayed maximum \citep{2006ApJ...649..939K, 2007ApJ...656..661K,2017ApJ...846...58H}. As a result, their colors are redder at maximum light, with $(B-V)_{Bmax}>0.5$ mag, and their $B-V$ color curves peak earlier ($\sim$15 days after $B$-band maximum) compared to normal SNe Ia ($\sim$30 days). Spectroscopically, 91bg-like SNe are characterized by prominent Ti II absorption features and strong O I $\lambda$7774 lines, indicative of low ionization and temperature. They also exhibit deeper Si II $\lambda$5972 absorption than normal SNe Ia, along with relatively low ejecta velocities, fast photometric decline, and an absence of high-velocity features. To first order, the luminosity of a SN depends on the mass of radioactive $^{56}$Ni synthesized in the explosion, with less luminous objects producing smaller amounts of $^{56}$Ni \citep{1982ApJ...253..785A,2000ApJ...530..744P}. Given their low luminosity, 91bg-like SNe synthesize relatively small amounts of radioactive $^{56}$Ni ($<0.1$ M$_{\odot}$) \citep{2006A&A...460..793S}.

In this paper, we present an analysis of the most complete sample of subluminous SNe Ia ever performed using the ZTF Data Release 2 (DR2; \citealt{2025A&A...694A...1R}, Smith et al in prep.), obtained between March 2018 and December 2020, and consisting of 3,628 spectroscopically confirmed SNe Ia. We focus on studying the photometric and spectroscopic particularities of different subluminous SN Ia subgroups, and determining the first homogeneous and well-defined sample of 91bg-like SNe Ia. 


\section{Sample selection}\label{sec:sample}

\begin{table}[t]
\centering
\caption{A summary of the numbers of SNe Ia of each subtype included in our initial subluminous sample and quality cuts applied in SALT2 and SNooPy to produce final sub-samples.}
\label{tab:class_numbers}
\begin{tabular}{lccc}
\hline
Subtype & All SNe & SALT2$^{a}$ & SNooPy$^{b}$\\
\hline\hline
{\it Gold} 91bg-like & 81 & 57 & 66 \\
{\it Silver} 91bg-like & 6 & 6  & 6 \\
86G-like & 12 & 12 & 12 \\
04gs-like & 18 & 18 & 18 \\
02es-like & 7 & 5 & 4 \\
\hline
Total  & 124 & 98 & 106 \\
\hline
\end{tabular}
\tablefoot{\small{ (a), (b) refer to the sub-samples after applying the cuts described in Section \ref{subsec:LC-fit}.}}
\end{table}

In this section we describe the selection of our subluminous SN Ia sample from the 3628 spectroscopically classified SNe Ia in the ZTF DR2, without applying any constrains or selection criteria (e.g. LC coverage, LC quality, redshift or phase), in order to increment our sample size and include a larger number of subluminous SNe Ia.

The spectral classification of the ZTF DR2 has been implemented using {\sc Typingapp}, a web-application tool specifically designed for this purpose, which allowed to assign a classification based on all available spectra with their first 30 best SNID fits \citep{2007ApJ...666.1024B}, together with the light-curve with a SALT2 \citep{2007A&A...466...11G} fit and light-curve parameters to verify the consistency of the spectral phase with the light-curve phase. The redshift of the SN Ia and its origin (e.g. catalogue, host galaxy features in the SN spectra, SNID best fit) was also available. Expert human assessment provided, when possible, subtype classifications to all SNe Ia released in the ZTF SNIa DR2 \footnote{\href{https://ztfcosmo.in2p3.fr/}{https://ztfcosmo.in2p3.fr/}}. In particular, those subtypes included ‘SNIa’, ‘SNIa-normal’, ‘SNIa-91bg’, ‘SNIa-91T’, and ‘SNIa-pec’. For more details about the Typingapp and its methodology of classification, see \cite{2025A&A...694A..10D}.

We initially select all 81 confirmed 91bg-like SNe Ia from the internal subclassification performed through \texttt{Typingapp}, and label them here as {\it Gold 91bg-like} candidates. In addition, we included six potential candidates that have 91bg-like SNe Ia characteristics, but that given the quality of either the spectra or the light-curve cannot be firmly classified as 91bg-like. We label them as {\it Silver 91bg-like} candidates. Additionally, we included in our sample other subtypes from the transitional region between normal and subluminous SNe Ia, classified in the ZTF DR2. This include 12 ‘1986G-like’, which are events known as transitional between ‘normal’ and 91bg-like SNe Ia \citep{1987PASP...99..592P}, 18 ‘04gs-like’ defined in \citealt{2025A&A...694A...9B} as a subclass lies between ‘normal’ and transitional 86G-like SNe Ia with stronger \ion{Si}{II} $\lambda$5972 than normal, but significantly weaker \ion{Ti}{II} features than 86G-like) and 7 ‘02es-like’ SNe Ia \citep{2012ApJ...751..142G} that show a slow light-curve evolution consistent with normal SNeIa, but with similar underluminous spectroscopic features. A summary of the number of SNe of each subtype included in our subluminous sample are listed in Table \ref{tab:class_numbers}.


\section{Methods}\label{sec:methods}

\subsection{Light-curve fitting}\label{subsec:LC-fit}

\subsubsection{\textit{SALT2}}\label{subsubsec:salt2}

We have obtained the Spectral Adaptive Light Curve Template for SN Ia (SALT2; \citealt{2007A&A...466...11G}) light-curve parameters for the full ZTF DR2 SNe Ia from the published tables of the ZTF DR2 \citep{2025A&A...694A...1R}. These include the peak magnitude in the $B$-band, the time of maximum in the $B$-band $t_{max}$, the $x1$ stretch parameter and the $c$ color parameter. To these, we implemented the following quality cuts: 
\begin{itemize}
\item $\sigma_{x1} < 1$;
\item $\sigma_{c} < 0.1$ mag;
\item $\sigma_{t_{max}} < 1$ days;
\item $lccoverage\_flag = 1$;
\item $fitquality\_flag =1$;
\end{itemize}
where lccoverage\_flag and fitquality\_flag define the good sampling and other basic cuts described in Table. 1 in \citet{2025A&A...694A...1R}
As a result of applying these cuts, we are left with 2,667 SNe Ia from the full ZTF DR2 sample. This includes 98 subluminous SNe Ia from our initial set of 124: 57 {\it Gold} 91bg-like, all six {\it Silver} 91bg-like, 12 86G-like, 18 04gs-like, and 5 of the 7 02es-like SNe Ia.

\subsubsection{\textit{SNooPy}}\label{subsubsec:snoopy}

We fit all 3628 SN Ia light-curves in the ZTF DR2 with the python-based package “Object-Oriented” for analysis of SN Ia light curves (SNooPy; \citealt{2011AJ....141...19B}). First, a spline interpolation is performed independently to each individual $gri$ light-curve, to serve as an initial guess for the SNooPy template-based fit. The first light-curve fit is done by using the {\it max\_model} with the k-correction measured from the updated \cite{2007ApJ...663.1187H} SN~Ia template. The fit provides an estimate of the time and magnitude at maximum for each band independently ($t_{\rm max}$ and $m_{\rm max}$), together with a stretch-color parameter $s_{BV}$ which is the time between B-band maximum light and the reddest point in the $B-V$ color curve normalized by 30 days, $s_{BV} = (t_{max}$/30 days). For those objects with $s_{BV}$ lower than 0.6, we repeated the fit estimating the k-corrections with the 91bg template instead included in SNooPy. Finally, for objects with a successful fit, we used the {\it EBV\_model2} that besides $s_{BV}$, $m_{\rm max}$, and $t_{\rm max}$, provides an estimate of the host galaxy extinction $E(B-V)_{\rm host}$. Similarly, we applied some quality cuts: 
\begin{itemize}
    \item $\sigma_{s_{BV}} < 0.1$;
    \item $\sigma_{EBVhost} < 0.1$ mag;
    \item $\sigma_{t_{max}} < 1$ days;
    \item $lccoverage\_flag = 1$ (from ZTF DR2 SALT2 table);
\end{itemize}
that reduced the sample to 2672 SNe Ia, including 81 of the 124 SNe Ia in our subluminous sample. 

\begin{figure}[t]
\centering
\includegraphics[width=\columnwidth]{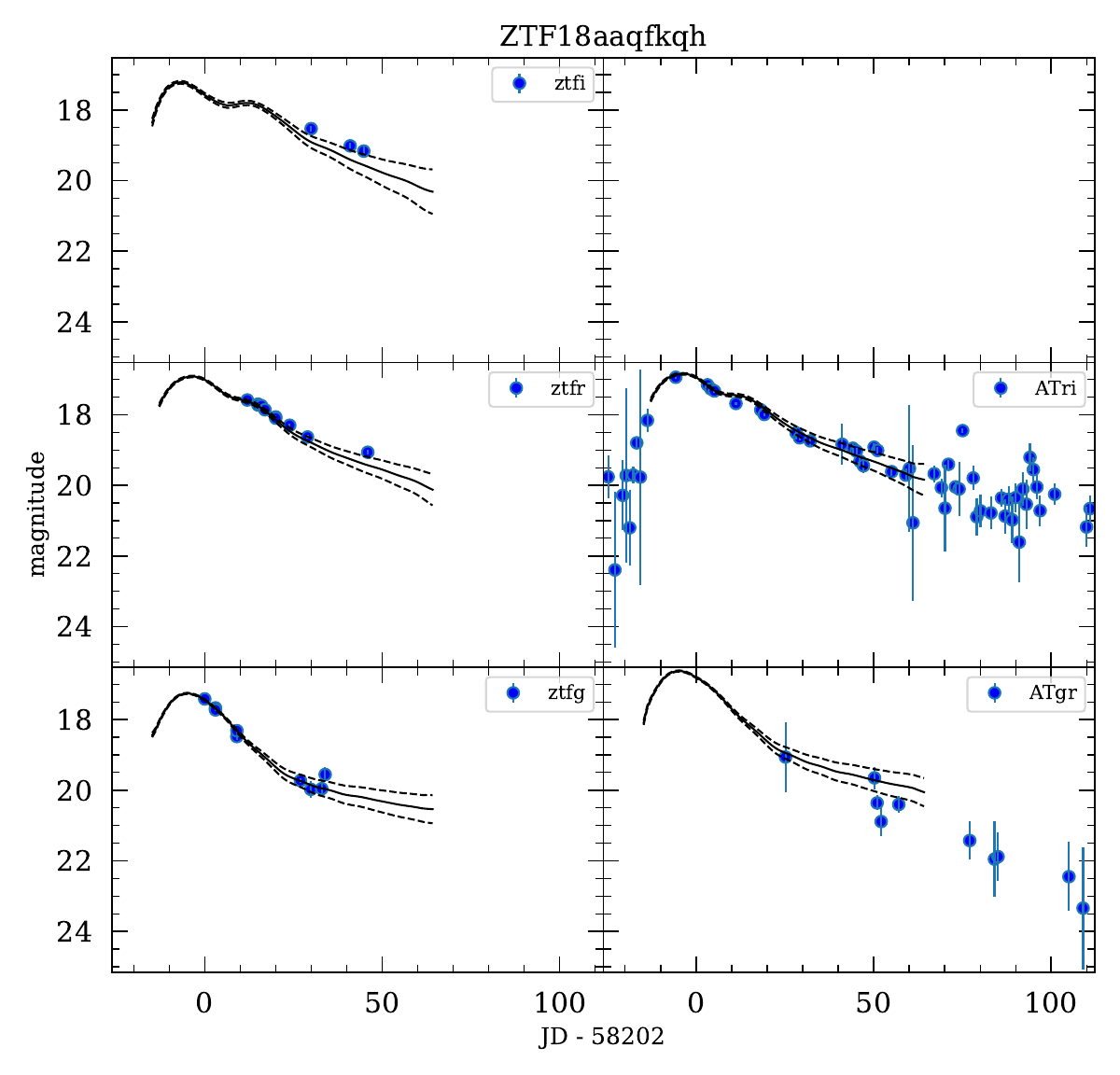}
\caption{An example of SNooPy fit of ZTF18aaqfkqh light-curves. The three ZTF $gri$ bands do not have observations before peak magnitude, making the determination of the time of maximum very uncertain. Including $co$ ATLAS data allows for a combined fit that adequately constrain the time of maximum and the light-curve width. \textit{Note} that the $ATgr$- and $ATri$-bands correspond to $c$- and $o$-ATLAS bands , respectively.}
\label{fig:ztf_vs_atlas}
\end{figure}

To increase the number of subluminous SNe Ia remaining for further analysis, we tried to recover some of those with bad SNooPy fits and for which we were not able to determine the time of maximum light due to the poor quality of their light curves, most lacking sufficient observational data around peak. To address this limitation, we supplemented ZTF light-curves with public $c-$ and $o-$band photometry obtained from the Asteroid Terrestrial-impact Last Alert System (ATLAS; \citealt{2018PASP..130f4505T}) forced photometry service\footnote{\href{https://fallingstar-data.com/forcedphot/}{https://fallingstar-data.com/forcedphot/}}. We repeated the SNooPy fits fitting ZTF and ATLAS data simultaneously. Figure \ref{fig:ztf_vs_atlas} shows an example of the simultaneous $grioc$ light-curve fit of ZTF18aaqfkqh. In this particular case, it is clear that ATLAS photometry before peak help constraining both the time of maximum and the stretch of the light-curve. We successfully improved or recovered light-curve parameters for up to 25 SNe Ia in our subluminous sample, increasing the number of well-fitted events to 106 out of a total of 124, split into the subtypes: 66 {\it Gold} 91bg-like, 6 {\it Silver} 91bg-like, 12 86G-like, 18 04gs-like, and 4 of the 7 02es-like SNe Ia. The measured parameters for our subluminous sample are presented in Table \ref{tab:LC_parameters}. For the remaining SNe Ia in the ZTF DR2, we include the results in the figures to serve as a comparison to our subluminous sample.

\subsection{Spectral characterization}\label{subsec:spex}

The ZTF DR2 SN Ia sample contains 3,628 spectroscopically confirmed SNe Ia, for which a total of 5,215 spectra were observed. From these, we selected 194 spectra our spectral analysis of the 124 subluminous SNe Ia in this study. 

\subsubsection{Velocities and pEWs}

\begin{figure}[t]
\centering
\includegraphics[width=\columnwidth]{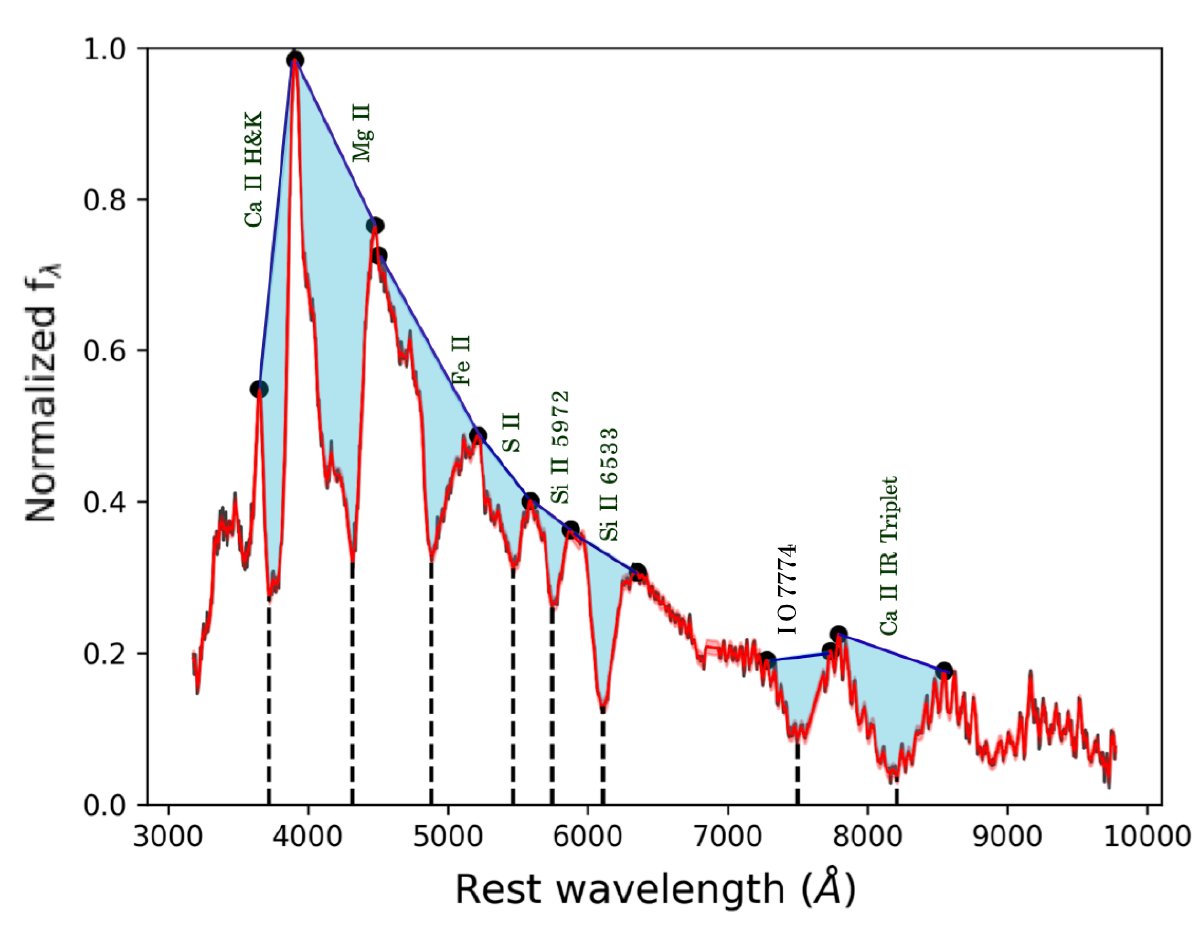}
\caption{Example spectrum of SN 2018aaz (ZTF18aabstmw) at 6 days before maximum light, highlighting the eight main spectral features. The corresponding pseudo-equivalent width (pEW) measurements are illustrated as blue-shaded regions.}
\label{fig:pew_spex}
\end{figure}

In order to study and analyze the spectral features such as velocities ($v$) and pseudo-equivalent widths ($pEW$), of the 194 spectra for our 124 subluminous SN~Ia of our sample, we have used a modified version\footnote{\href{https://github.com/anthonyburrow/spextractor}{https://github.com/anthonyburrow/spextractor}} of the code \texttt{Spextractor}\footnote{\href{https://github.com/astrobarn/spextractor}{https://github.com/astrobarn/spextractor}} \citep{2019PhDT.......134P,2020ApJ...901..154B}. This modified version has introduced the down-sampling of spectral information, in order to reduce computational cost. Additionally, adjustments were implemented to enhance the accuracy of the GP model and smoothing techniques for a given spectrum. As demonstrated in \cite{2020ApJ...901..154B}, these improvements aim to ensure the model yields more reliable measurements of the parameters. As illustrated in Fig.~\ref{fig:pew_spex}, we show an example of a SN Ia spectrum with the $pEW$s of the identified spectral features along the spectrum, represented as follows: \ion{Ca}{II} IR triplet, \ion{O}{I} $\lambda$7774, \ion{Si}{II} $\lambda$6355, \ion{Si}{II} $\lambda$5792, \ion{S}{II}, \ion{Fe}{II}, \ion{Mg}{II}, and \ion{Ca}{II} H\&K.

\begin{figure}[t]
\centering
\includegraphics[width=\columnwidth]{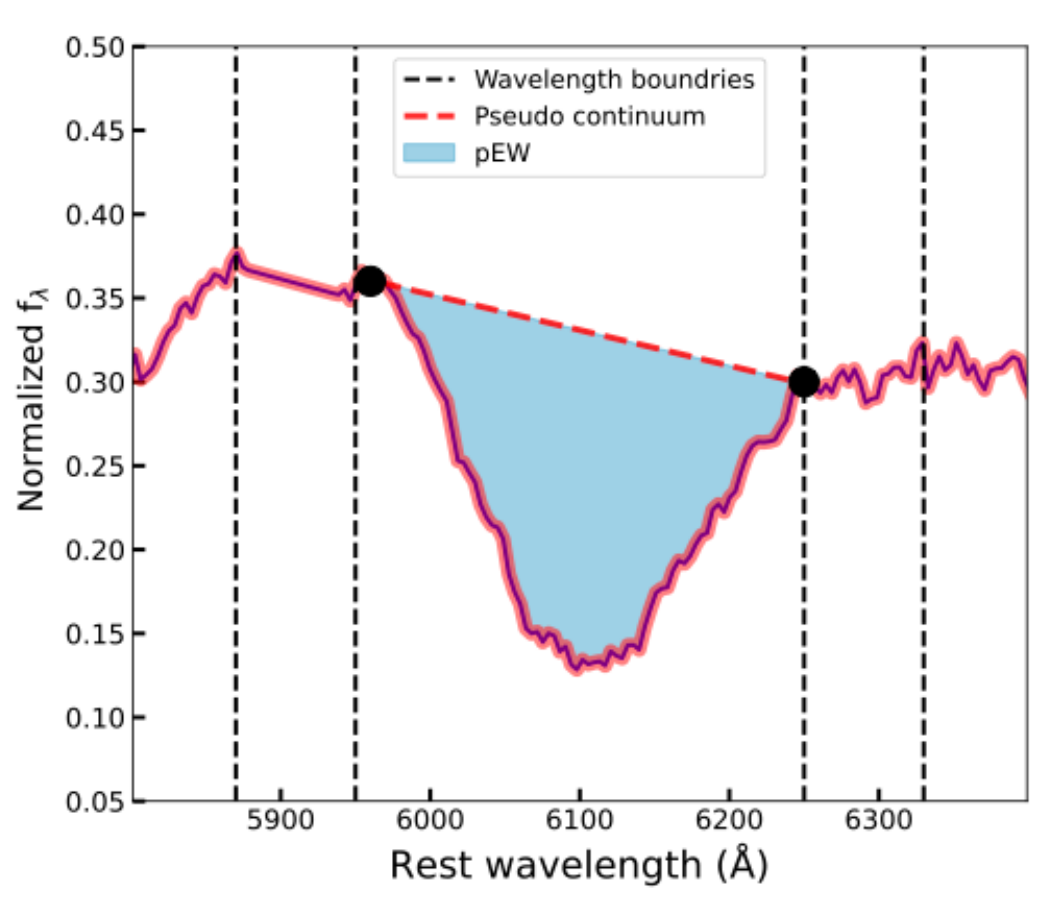}
\caption{Demonstration of a measurement of a pseudo-equivalent width (pEW) for the \ion{Si}{II} $\lambda$6355 feature with \texttt{Spextractor}. The rest-framed spectrum with its uncertainty is shown in red. The vertical dashed black lines determine the wavelength boundaries of the feature. A dashed red line represents the pseudo-continuum connecting both sides of the spectral feature, and the shaded blue area is the integral between the pseudo continuum and the binned spectrum within the defined wavelength boundaries, which defines the pEW.}
\label{fig:pew}
\end{figure}

The velocities are calculated by identifying the minimum of the smoothed spectral features in observed wavelength relative to its rest wavelength, and by converting these wavelength shifts into velocities using the relativistic Doppler formula \citep{2006AJ....131.1648B} (as illustrated in Fig. \ref{fig:pew_spex}). The definitions of the spectral features and their wavelength boundaries are taken from  \citet{2013ApJ...773...53F}. For the $pEW$s, the code sets a {\it pseudo-continuum} by fitting a straight line to the peak normalized flux on either side of a spectral feature as demonstrated in Fig.~\ref{fig:pew}. To determine the $pEW$s, we select wavelength boundaries of each observed feature. These boundaries are chosen manually following a visual examination of the rest-framed spectrum. Finally, the pseudo-continuum is drawn as a straight line connecting the points where the spectrum intersects each boundary. The $pEW$ represents the shaded blue area, calculated as the integral between the pseudo continuum and the binned spectrum within the defined wavelength boundaries. 

\subsubsection{Spectral phases}

\begin{figure*}[t]
\centering
\includegraphics[width=0.9\textwidth]{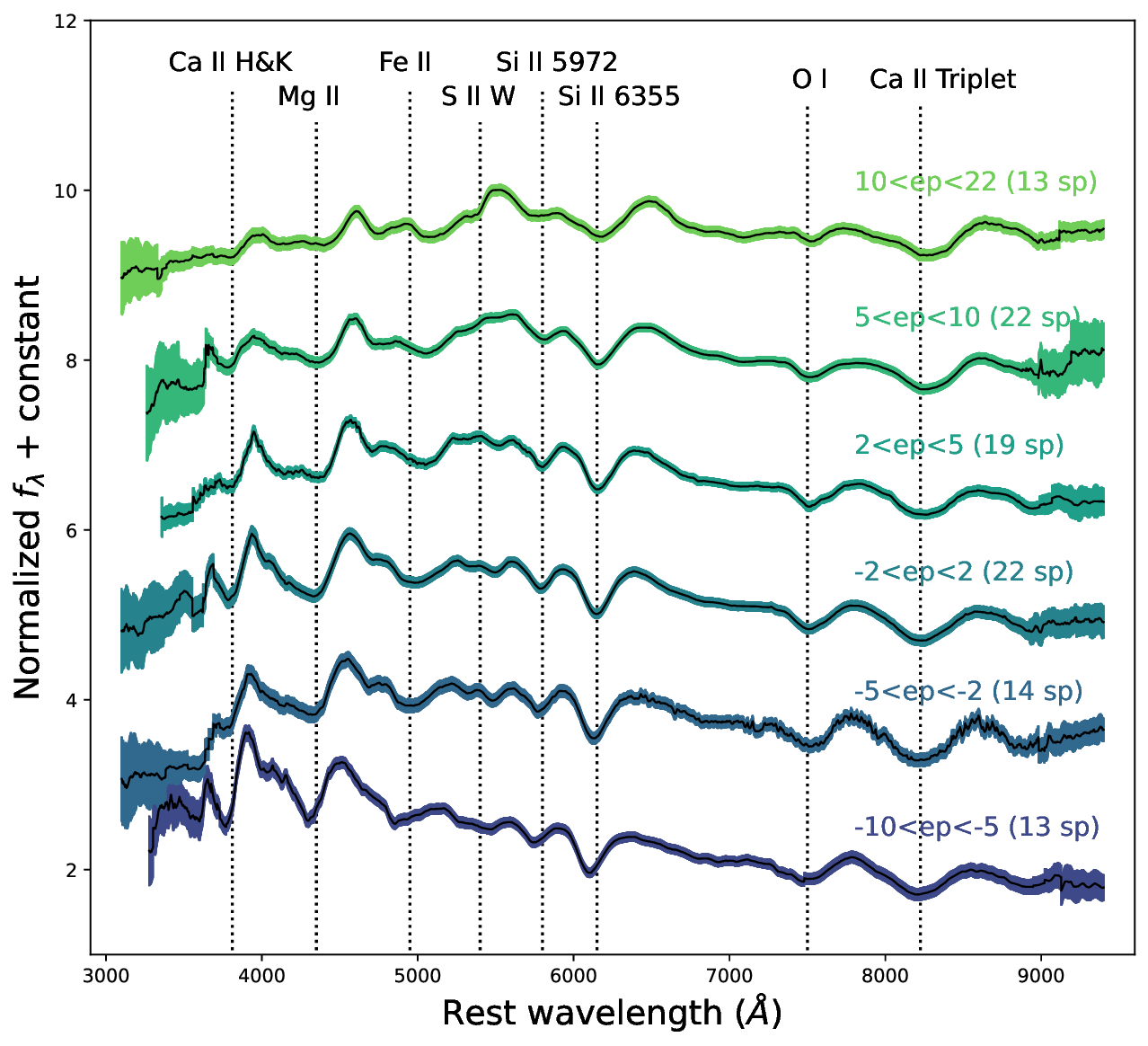}
\caption{Spectral averages across six phase bins for all 103 91bg-like SNe Ia spectra in our sample. Key spectral features are marked with vertical lines at their minimum in the spectrum at maximum light, allowing their velocity evolution to be tracked across different phases.}
\label{fig:specaverage1}
\end{figure*}

We recalculated the phases of all the available spectra of our subluminous SN Ia sample, using the spectra observation epoch, the final estimated values of maximum time $t_{max}$ at $B$-band from the SNooPy fits, and the SN redshift $z$, by doing $phase = \left( t_{obs} - t_{max}\right)/(1 + z)$. Most of the spectral diagnostics for SN~Ia (e.g the \citealt{2006PASP..118..560B} diagram) need the estimate of the spectral parameters at or around maximum light. Therefore, for further analysis we only kept those ejecta velocities ($v$) and pseudo-equivalent widths ($pEW$) measurements from spectra with phases between $-5<{\rm ep}<5$ days. In case that a SN has spectra outside that range, that SN is not included in our spectral analysis. For SNe with more that one spectrum within that range, we applied the following criteria: 
\begin{itemize}
\item If all spectra are at positive or negative phases, we keep the one closest to maximum light.
\item If there are spectra at positive and negative phases, we interpolate the spectral parameters to obtain a weighted measurement at the epoch of maximum light (0 days).
\end{itemize}
Thus, by applying these criteria, we have a final set of spectral parameters for 89 subluminous SNe Ia in our sample: 55 ‘91bg-like’, 12 ‘86G-like’, 18 ‘04gs-like’ and 4 ‘02es-like’. In Table \ref{tab:vel-pew}, we present the values, including \ion{Si}{II} $\lambda$6355 velocity and pseudo-equivalent widths of spectroscopic features (\ion{Si}{II} $\lambda$6355, \ion{Si}{II} $\lambda$5972, \ion{Mg}{II} $\lambda$4000, \ion{O}{I} $\lambda$7774, and \ion{Ca}{II} NIR) of all 124 SNe in the initial sample, together with the spectral phase and spectroscopic classification of each SN.

\subsubsection{Spectral averages}

The ZTF DR2 dataset represents the first large, complete, and unbiased sample of SNe Ia up to  $z\sim0.06$, discovered using a single telescope. This offers the distinct advantage of enabling more precise determinations of the ratios between different SN Ia subtypes, which is crucial for measuring quantities such as event rates. For a specific subtype, such as 91bg-like SNe Ia in our case, a complete sample should represent the population’s full diversity of properties. So, using the available SN spectra, we construct averages in different phase bins to provide a global spectroscopic characterization of this subluminous SN Ia subtype. The six bins, each representing a different phase range in rest-frame days, are defined as follows: (-10 to -5), (-5 to -2), (-2 to +2), (+2 to +5), (+5 to +10), and (+10 to +22). 

We processed a total of 103 spectra from our sample of 91bg-like SNe Ia. After correcting for Milky Way (MW) extinction using dust maps from \cite{2011ApJ...737..103S} with a \cite{1999PASP..111...63F} extinction law with the average galactic $R_V=3.1$, and shifting the spectra to the rest-frame, we smoothed the spectra using a 1D Savitzky-Golay filter. The polynomial order of the filter was adjusted based on the spectral resolution, ranging from order 11 for the lowest resolution (SEDMachine) to order 31 for the highest resolution (Keck). During this process, the spectra were resampled to a uniform wavelength step of 1\AA. When averaging the spectra, they were weighted taking into account the signal-to-noise ratio, the spectral resolution, and a gaussian wavelength dependent component to account for vignetting in the detector (this gives more weight to the central part of the spectrum and less to the two blue and red ends). The resulting average spectra for 91bg-like SNe Ia are illustrated in Fig. \ref{fig:specaverage1}, where the spectral evolution of these spectra and their features are identified prior to maximum light, at maximum light, and posterior to maximum.

This process was repeated also including in the first step a host galaxy extinction correction from the SNooPy $E(B-V)_{\rm host}$ parameter, assuming a \cite{1999PASP..111...63F} extinction law but with an $R_V=2.5$, more suited for extragalactic dust as shown in multiple SN~Ia studies (e.g. \citealt{2014ApJ...789...32B}). Similarly, we repeated the process for other subtypes; 86G-like and 04gs-like, although due to the reduced number of spectra (15 and 25, respectively) the only reliable average spectra are those around maximum light. Also, we note that the paucity of 02es-like events in our sample precludes the construction of a meaningful averaged spectrum for this subtype.

\subsection{Host galaxies}\label{subsec:hostGal}

Host galaxy association as well as measurements of mass and rest-frame $g-z$ colors is described in the ZTF DR2 release \citep{2025A&A...694A...1R}. These properties were derived from stellar population synthesis (SPS) applied to Pan-STARRS $grizy$ photometry, measured using {\sc HOSTPHOT} \citep{2022JOSS....7.4508M}. These measurements include both global properties, obtained from elliptical apertures encompassing the entire galaxy, and local properties, measured within circular apertures of 2 kpc radius centered on the SN position. Additionally, ZTF DR2 provides directional light radius distances ($d_{DLR}$; \citealt{2006ApJ...648..868S}; \citealt{2016AJ....152..154G}), which are defined as the distance to the SN from the galaxy core, normalized by the distance from the galaxy center to the elliptical aperture radius in the direction of the SN. We obtain these two host galaxy parameters, both global and local, and the $d_{DLR}$ to study particularities for the subluminous SN Ia sample.


\section{Results}\label{sec:result}

\subsection{Light-curve parameters}\label{subsec:lcpars}

\subsubsection{Time of maximum light}\label{subsec:tmax}

\begin{figure}[t]
  \centering
\includegraphics[width=\columnwidth]{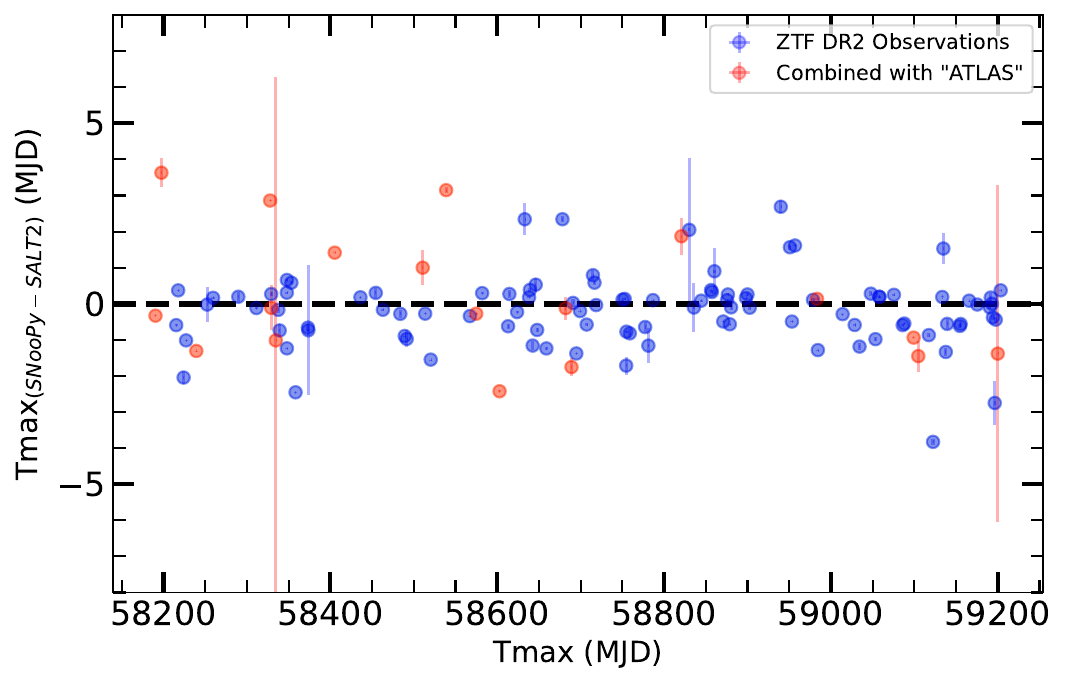}
\caption{SN time of maximum light residuals comparison between SNooPy and SALT2 light-curve models. Blue points indicate differences of time of maxima when fitted to ZTF data only, and red dots indicate differences when ATLAS data is included in the SNooPy fits only.}
\label{fig:tmax}
\end{figure}

The times of maximum light derived from the three fitting methods are compared here to identify the most consistent and accurate estimates. Figure \ref{fig:tmax} shows the differences in the time of maximum light obtained with SNooPy —using only ZTF data (blue) and with the inclusion of ATLAS photometry (red)— relative to those reported by SALT2 in ZTF DR2. When using only ZTF data, the differences between SNooPy and SALT2 show a scatter of $\sigma = 0.96$ days. This scatter may be explained by the fact that SALT2 is not fully optimized or trained for fitting light curves of subluminous SNe Ia, whereas SNooPy is. When ATLAS data are incorporated into the SNooPy fits, the estimated times of maximum light become more precise in some cases. This improvement is due to the additional early-time data provided by ATLAS, particularly around or before peak brightness, which are often missing from ZTF alone. As expected, the inclusion of ATLAS data leads to larger scatter in the comparison ($\sigma({\rm SNooPy - SALT2}) = 1.73$ days). These increased values reflect that some of the original time-of-maximum estimates —based solely on ZTF— were slightly inaccurate. For the remainder of this paper, we adopt the SNooPy times of maximum derived from the combined ZTF+ATLAS fits.

\subsubsection{Distributions of light-curve parameters}

\begin{figure*}[t]
\centering
\includegraphics[trim=0.2cm 0.3cm 0.2cm 0.2cm,clip=True,width=\columnwidth]{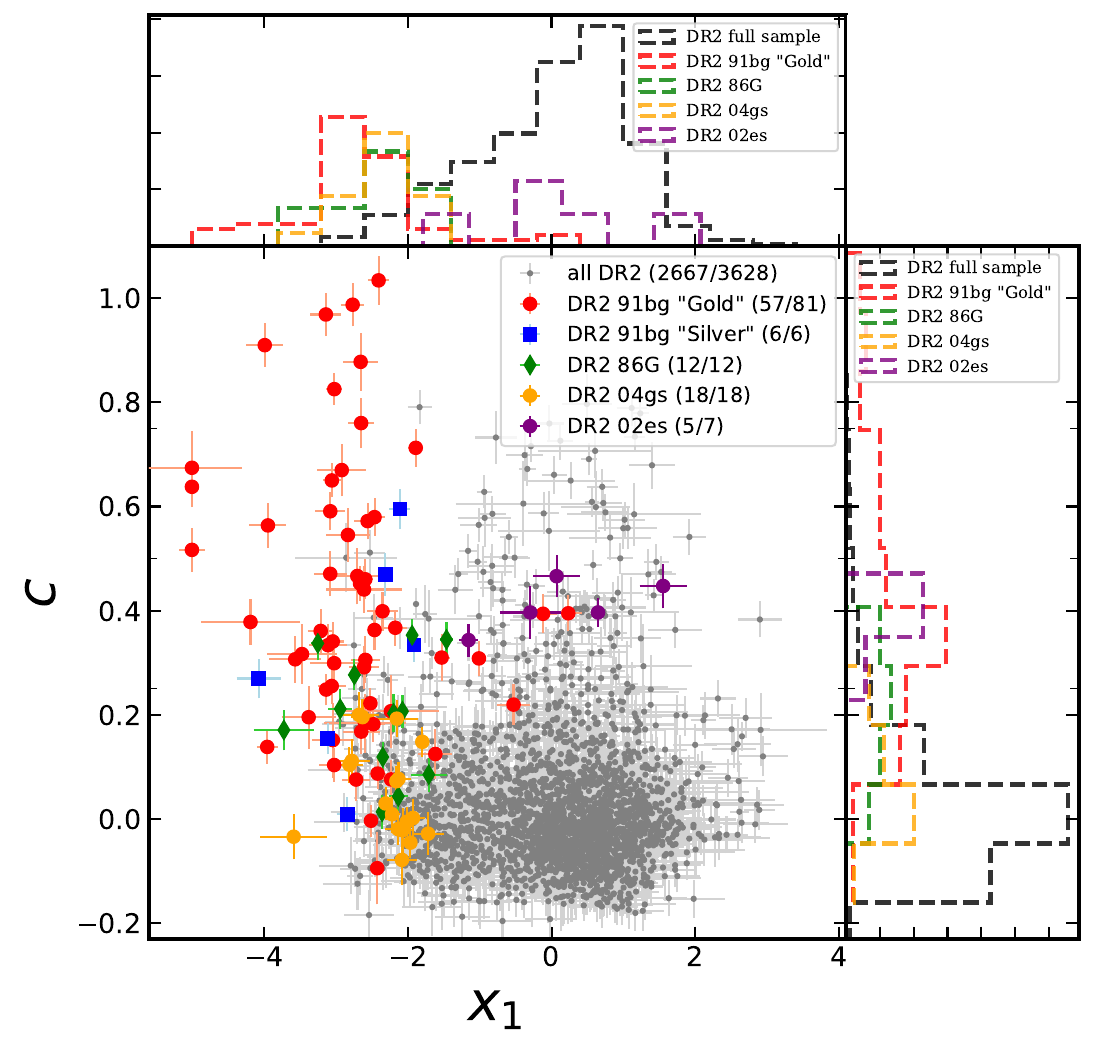}
\hspace{2mm}
\includegraphics[trim=0.2cm 0.3cm 0.2cm 0.2cm,clip=True,width=\columnwidth]{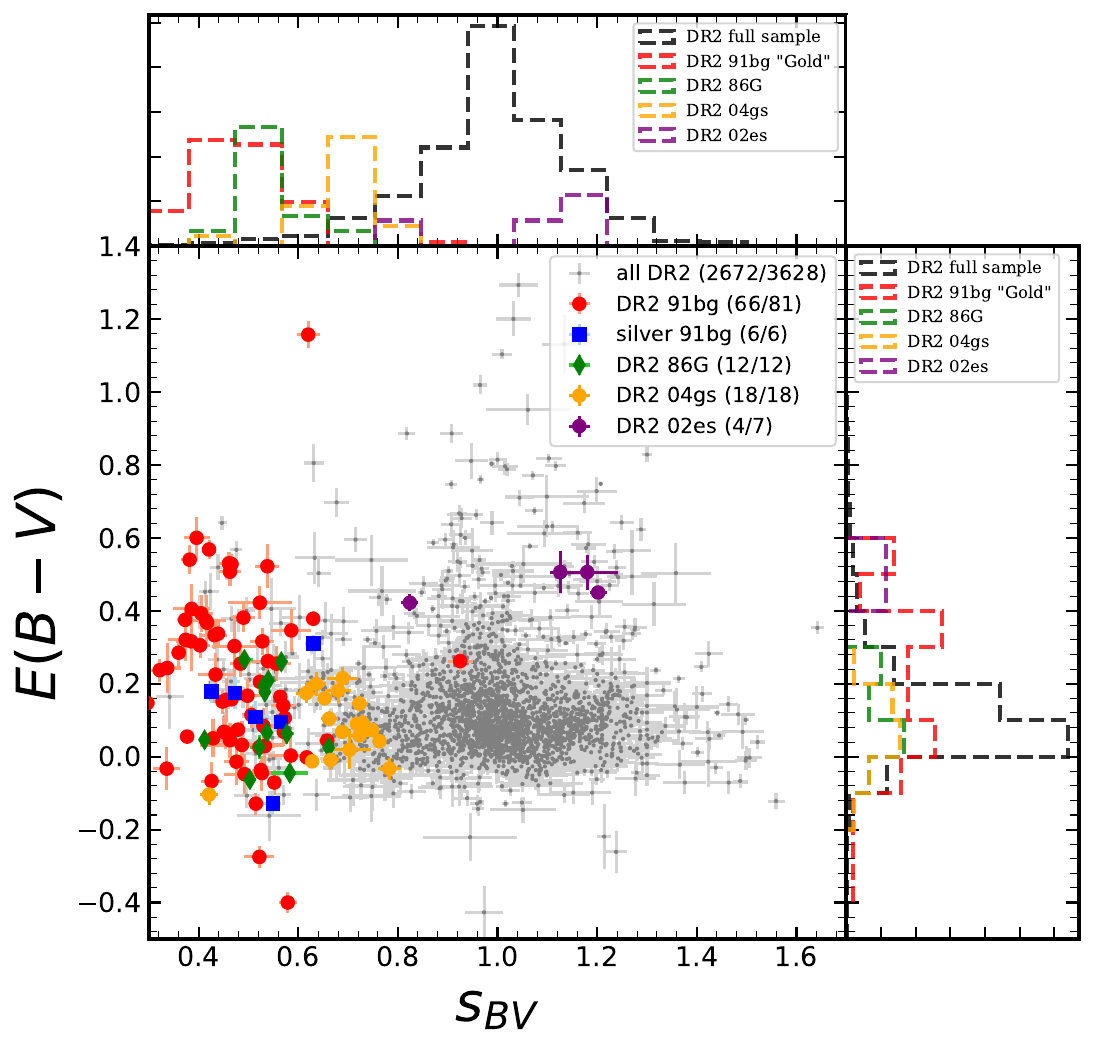}
\caption{Left: The SALT2 $x1$ vs. $c$ diagram, along with the distributions of these parameters, is shown for all normal SNe Ia in ZTF DR2 and the subluminous groups: 91bg-like "Gold," 91bg-like "Silver," 86G-like, 04gs-like, and 02es-like. The numbers in the legend represent only the successfully fitted light curves from SALT2. Right: Similarly, the SNooPy $s_{BV}$ vs. $E(B-V)$ diagram, along with the distributions of the parameters, for all normal SNe Ia in ZTF DR2 and the subluminous groups. The numbers in the legend represent the successful light-curve fits from SNooPy.}
\label{fig:color}
\end{figure*}

The left panel of Fig. \ref{fig:color} shows the SALT2 stretch parameter $x1$ versus the color parameter $c$, including histograms along each axis, for both subluminous SNe Ia from our sample and normal SNe Ia from the full ZTF DR2 sample. The legend indicates the number of good fits for each subtype. Normal SNe Ia are broadly distributed along the x-axis, with $-3 \leq x1 \leq 3$, whereas subluminous SNe Ia are confined to a narrower range of $-4 \leq x1 \leq -2$. Along the y-axis, the color parameter $c$ tends to be higher (redder) for subluminous SNe Ia compared to normal ones — especially for 91bg-like SNe, which can reach values up to $c \sim 1.0$.

\begin{figure*}[t]
\centering
\includegraphics[trim=11.5cm 2cm 13cm 7cm,clip=True,width=\textwidth]{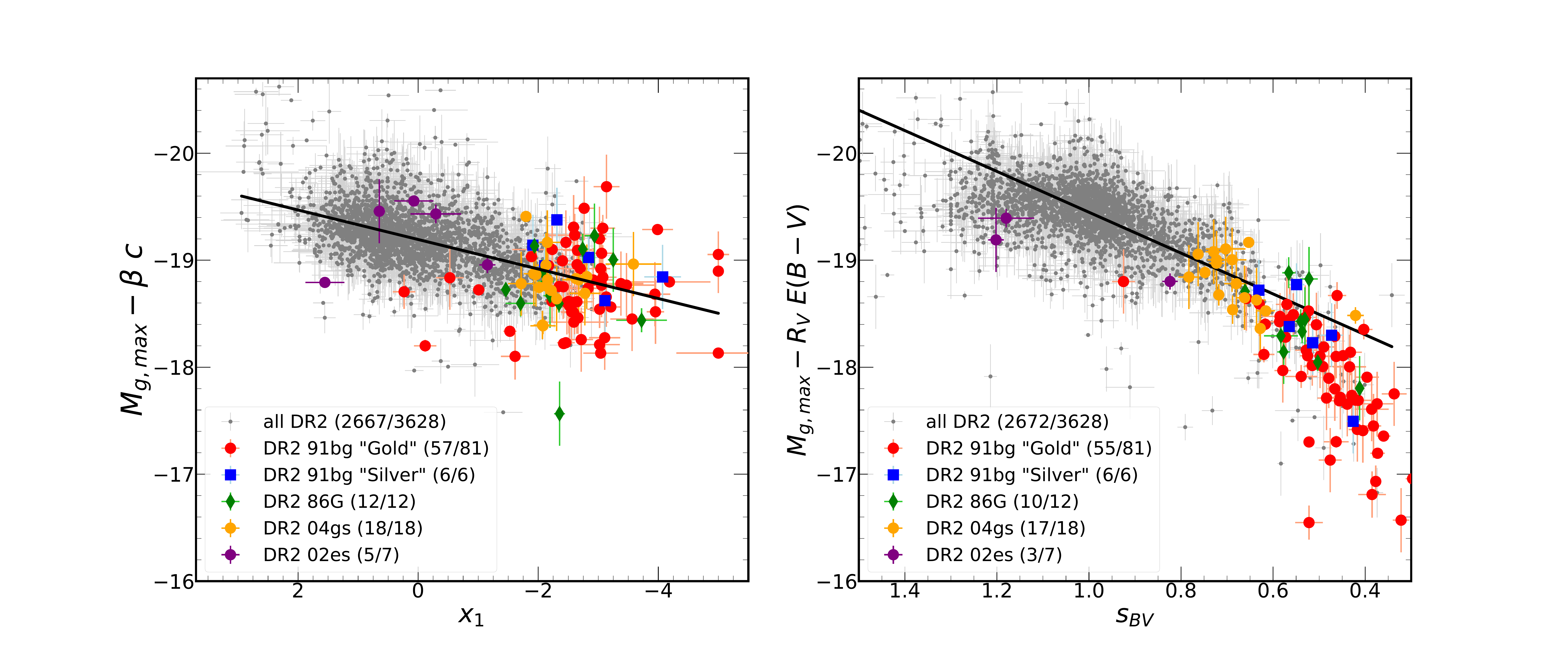}
\caption{The absolute peak magnitudes in the $g$-band of our subluminous sample (in different colors) and the full ZTF DR2 in the background (in grey) plotted against the SALT2 stretch parameter $x1$ (left Panel) and the color stretch $s_{BV}$ (right Panel). These magnitudes are corrected for host galaxy extinction in terms of $R_V E(B-V)$ and from intrinsic color differences in terms of $\beta c$, where  $R_V= 2.31$ and  $\beta=3.31$. The black lines represent the 1st-order polynomial best fits to the normal SNe Ia for ($0.4<s_{BV}<1.4$) and ($-5<x1<3$). The numbers shown in the legends represent the numbers of good fits obtained from the subluminous and DR2 full sample.}
\label{fig:phillips}
\end{figure*}

Similarly, the right panel of Figure \ref{fig:color} presents the distributions of the color stretch parameter $s_{BV}$ and the color excess $E(B-V)$ derived using SNooPy, along with histograms along the top and right axes. Subluminous SNe Ia show $s_{BV}$ values between $0.3 \leq s_{BV} \leq 0.8$, while normal SNe Ia range from $0.6 \leq s_{BV} \leq 1.4$. The distributions of $E(B-V)$ for both subluminous and normal SNe Ia fall within $-0.3 \leq E(B-V) \leq 0.8$. In both panels, subluminous SNe Ia are generally associated with narrower, faster-evolving light curves. The exception is the 02es-like SNe Ia, which display slower and broader light-curve evolution, as reflected in their higher values of both $s_{BV}$ ($\geq 1$) and $x1$ ($\geq 0$).

Unlike the SALT2 color parameter $c$, which captures both intrinsic and extrinsic color variations, the SNooPy $E(B-V)$ parameter primarily reflects extrinsic effects, such as host galaxy extinction. Intrinsic color differences in SNooPy are accounted for by the $s_{BV}$ parameter. As shown in Fig. \ref{fig:color}, subluminous SNe Ia generally exhibit a broader color distribution than normal SNe Ia, especially the 91bg-like subtype, which reaches higher color values. However, the histograms suggest that differences in $E(B-V)$ are not as pronounced. This may indicate that the redder colors of subluminous SNe Ia are largely due to intrinsic differences (see, e.g., \citealt{2025A&A...694A..10D}).

\subsubsection{Peak magnitude vs. light-curve width and color diagrams}

We investigate the empirical relation between SN~Ia peak luminosity and light-curve width by plotting the color- and extinction-corrected $g$-band peak absolute magnitudes of SNe Ia against the SALT2 stretch parameter $x1$ and the SNooPy color-stretch parameter $s_{BV}$. In the left panel of Fig. \ref{fig:phillips}, we present the SALT2 results corrected for color differences using a color law of the form $\beta c$, with $\beta = 3.31$, as determined by \cite{2025A&A...694A...4G}. After this correction, the peak magnitudes exhibit a linear relation with the $x1$ parameter across both normal and subluminous SN Ia populations. Subluminous SNe Ia, which would otherwise appear fainter, are effectively brought onto the same linear relation by this correction, extending the trend towards lower $x1$ values. 

In the right panel of Fig. \ref{fig:phillips}, we show the peak magnitudes corrected for extinction using the term $R_V E(B-V)$, where $R_V$=2.31, which is the equivalent value of that used for SALT2 ($\beta \approx R_B = R_V +1$). In this case, subluminous SNe Ia deviate from the linear relation followed by normal SNe Ia, as $E(B-V)$ only accounts for extrinsic extinction, not intrinsic color differences. This distinction becomes clear when a linear fit to only the normal SNe Ia is extrapolated toward lower $s_{BV}$ values: subluminous SNe Ia fall below this trend, indicating they are intrinsically redder. The 02es-like SNe Ia also lie below the linear relation but have higher $s_{BV}$ values than other subluminous SNe Ia. This suggests that $s_{BV}$ does not fully capture their peculiarities, whereas the SALT2 color parameter $c$ does. These SNe are characterized by broader light curves but fainter peak magnitudes, which are likely due to their cooler, low-ionization ejecta. While their slow light-curve evolution resembles that of normal SNe Ia, they share several spectroscopic features with subluminous SNe Ia \citep{2025A&A...694A..10D}.

\subsection{Spectral parameters}\label{subsec:specpara}

\subsubsection{Branch classification diagram}

\begin{figure}[t]
\centering
\includegraphics[trim=1cm 0.8cm 1.6cm 2.2cm,clip=True,width=\columnwidth]{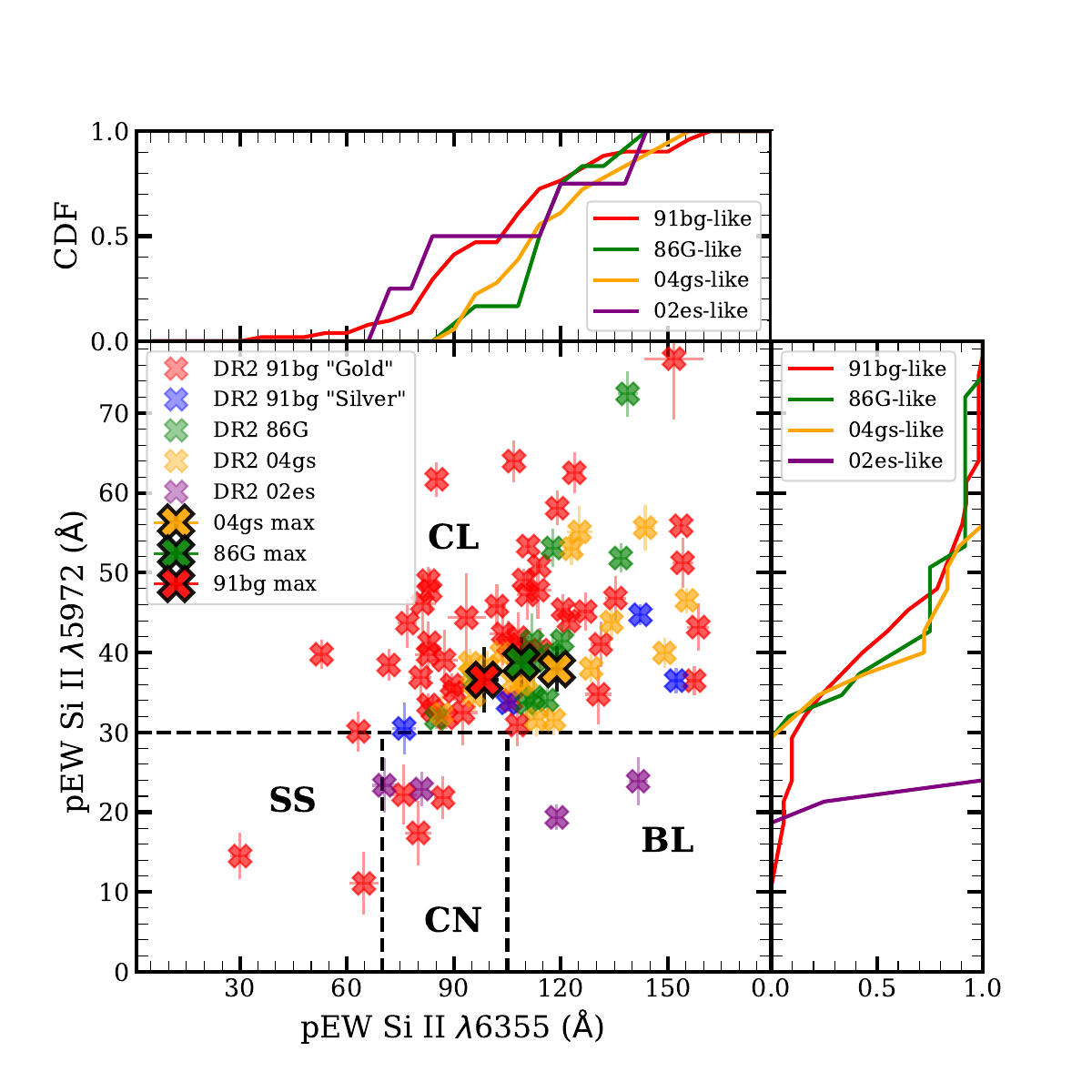}
\caption{Branch classification diagram plotted as the pEW of \ion{Si}{II} $\lambda$6355 against the pEW of \ion{Si}{II} $\lambda$5792 for the subtypes of our subluminous SNe Ia (91bg-like, 86G-like, 04gs-like and 02es-like) with spectra between $-$5 and +5 days with respect to maximum light, showing the spectral classification into 4 subclasses on the diagram: ‘core normal’ (CN), ‘shallow silicon’ (SS), ‘broad line’ (BL), and ‘cool’ (CL). Additionally, the calculated spectral averages for the three subtypes (91bg-like, 86G-like and 04gs-like) at epoch of maximum light ($-$2 < d < +2) are plotted on the Branch Diagram and the distributions of the pEWs for all subtypes are presented as shown in the legend.}
\label{fig:branchdiagram}
\end{figure}

Figure \ref{fig:branchdiagram} presents the Branch classification diagram \citep{2006PASP..118..560B}, which is based on the $pEW$s of the \ion{Si}{II} $\lambda$6355 and \ion{Si}{II} $\lambda$5792 spectral features. This diagram is widely used in the literature to categorize SNe Ia into four spectral subtypes: core normal (CN), shallow silicon (SS), broad line (BL), and cool (CL). We include the distributions of the measured pEWs for both \ion{Si}{II} $\lambda$5792 and \ion{Si}{II} $\lambda$6355 lines for all subtypes in our subluminous sample, shown along the top and right margins of the figure. Out of 89 SNe Ia in our sample —comprising 51 91bg-like Gold, 4 91bg-like Silver, 12 86G-like, 18 04gs-like, and 4 02es-like— with spectra taken within $\pm$5 days of maximum light, approximately 90\% (80 SNe) fall within the CL region, as expected for subluminous SNe Ia. Of the 9 SNe that fall outside the CL region, four are 02es-like. Although these exhibit some spectral similarities to 91bg-like SNe Ia, they generally show weaker \ion{Si}{II} $\lambda$5792 absorption.

\begin{figure*}[!t]
\centering
\includegraphics[width=0.8\textwidth]{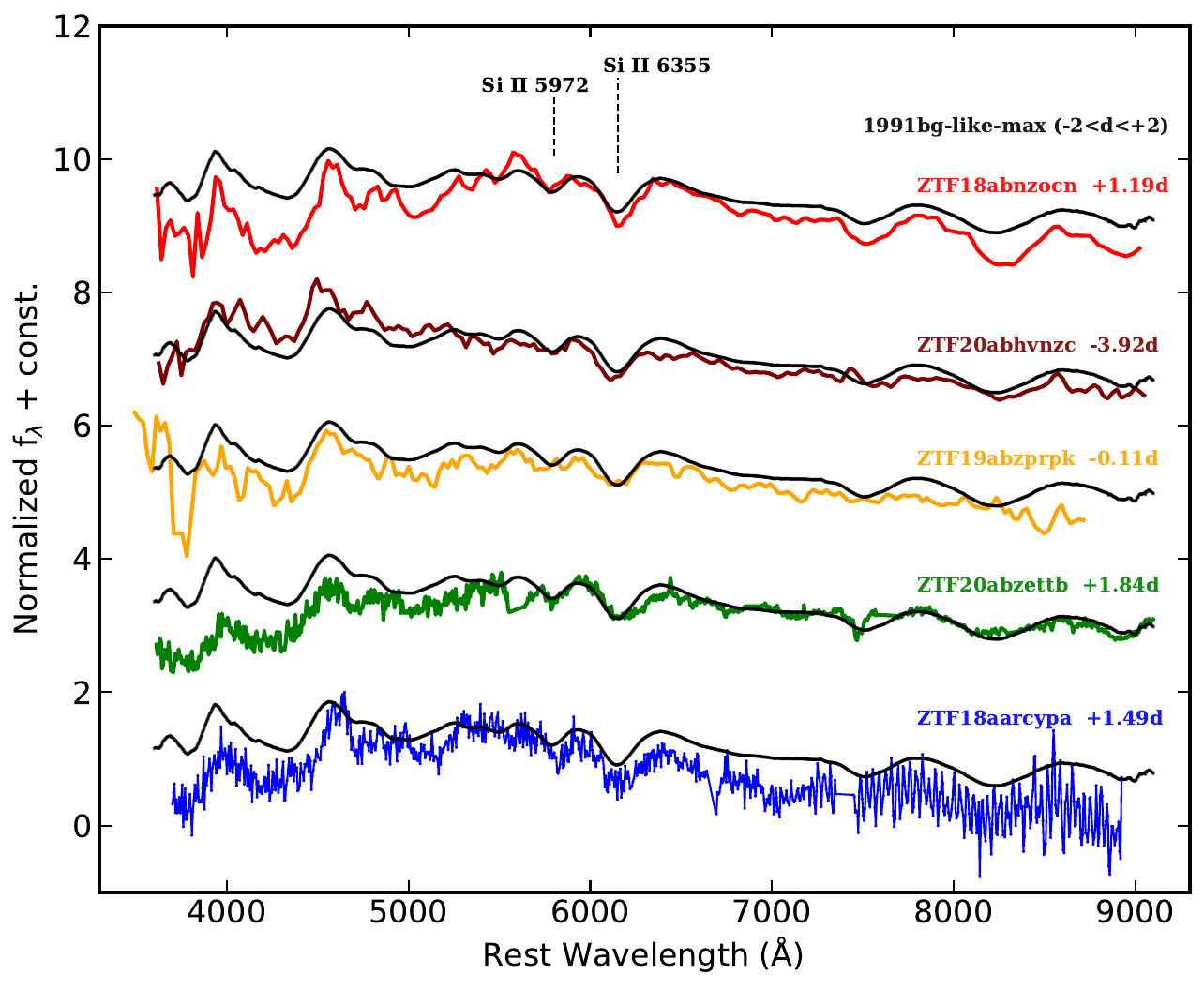}
\caption{The spectral average of the 91bg-like SN Ia (black) at epoch of maximum light ($-$2 < d < +2) compared to the spectra of the five 91bg-like SNe Ia identified on the Branch diagram (see, Fig. \ref{fig:branchdiagram}) in the lower regions with low pEW values, where the two \ion{Si}{II} $\lambda$5972 and $\lambda$6355 spectral features of these spectra are clearly affected by host galaxy contamination and/or low spectral resolution from the instrument causing inaccurate measurements of these absorption lines.}
\label{fig:specoutlier}
\end{figure*}

The remaining five SNe that fall outside the CL region are classified as 91bg-like {\it Gold}, but exhibit significantly lower \ion{Si}{II} $\lambda$5792 pEW values. We compare their spectra to the average 91bg-like SN Ia spectrum constructed at phases $-$2 < d < +2 days from maximum light. As shown in Fig. \ref{fig:specoutlier}, the \ion{Si}{II} $\lambda$5792 and \ion{Si}{II} $\lambda$6355 absorption features in these individual spectra appear notably weak. This could be due to host galaxy contamination or a low signal-to-noise ratio in the observations. Four of these spectra—ZTF18aarcypa, ZTF18abnzocn, ZTF19abzprpk, and ZTF20abhvnzc—were obtained using the SED-Machine instrument mounted on the 60-inch telescope (P60; 1.5m) at Palomar Observatory. The fifth, ZTF20abzettb, was observed with the larger 200-inch Hale Telescope (P200; 5m) at the same observatory. ZTF20abzettb exploded about 2 arcsec from the galaxy core and, despite being observed with a larger aperture telescope, its spectrum still shows evidence of contamination from host galaxy light due to limited spatial resolution. While the spectrum displays prominent \ion{Si}{II} $\lambda$5792 and broad, deep \ion{Ti}{II} absorption around $\sim$4000 \AA — features characteristic of 91bg-like SNe — the host galaxy continuum likely contributes significantly to the observed flux, artificially reducing the normalized $pEW$ values. The four SEDMachine-P60 spectra do exhibit signs of the \ion{Ti}{II} feature and the \ion{Si}{II} $\lambda$5792 absorption, but the data are notably noisy. 

Upon examining the locations within the host galaxies, we find that three out of the four SNe (excluding ZTF18abnzocn) occurred near the galactic core. This suggests that the combination of a small-aperture telescope and strong background contamination may explain the low pseudo-equivalent width ($pEW$) values observed for the \ion{Si}{II} $\lambda$5792 line. This interpretation is supported by the distances normalized by the directional light radius ($d_{DLR}$) obtained from ZTF DR2: all four spectra (ZTF18aarcypa, ZTF20abzettb, ZTF19abzprpk, and ZTF20abhvnzc) have $d_{DLR} < 0.2$. Such low $d_{DLR}$ values are typically associated with regions of high local surface brightness, indicating significant contamination from host galaxy light. This contamination can lead to artificially reduced measurements of the \ion{Si}{II} $\lambda$6355 pEW, as discussed in \citet{2025A&A...694A..13B}. In contrast, ZTF18abnzocn has a higher $d_{DLR}$ value of 3.13. Although the \ion{Si}{II} $\lambda$5792 line appears weak, its spectrum exhibits other key 91bg-like features, such as broad \ion{Ti}{II} and strong \ion{O}{I} and \ion{Ca}{II} absorptions. Moreover, an earlier spectrum taken at $-9$ days resembles that of a typical 91bg-like SN. The light curve also lacks a secondary maximum, and the SALT2 color parameter is notably high ($c = 0.97$ mag). These characteristics suggest that the weak \ion{Si}{II} $\lambda$5792 feature in this case may be due to suboptimal data reduction or poor observing conditions.

\begin{figure*}[h]
\centering
\includegraphics[width=\textwidth]{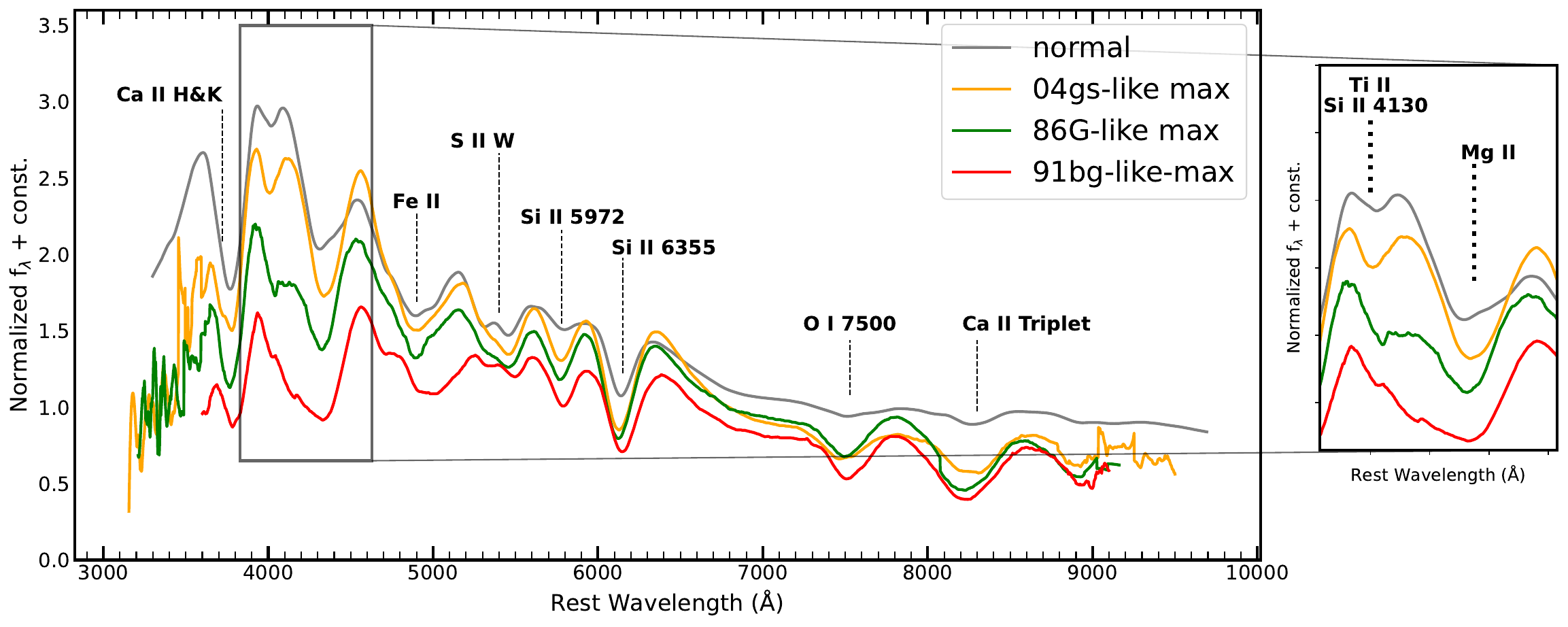}
\caption{Constructed average spectra of 04gs-like, 86G-like, 91bg-like SNe Ia at the epoch of maximum light (-2 < d < +2), alongside with the spectrum of the normal SN~2011fe at peak. A zoom-in view of this region highlights the varying strength of the \ion{Ti}{II} absorption feature across the different subtypes.}
\label{fig:specaverage3}
\end{figure*}

\begin{figure}[t]
\centering
\hspace{-0.7cm}
\includegraphics[trim=0.3cm 0cm 0cm 0cm, clip=True,width=1.05\columnwidth]{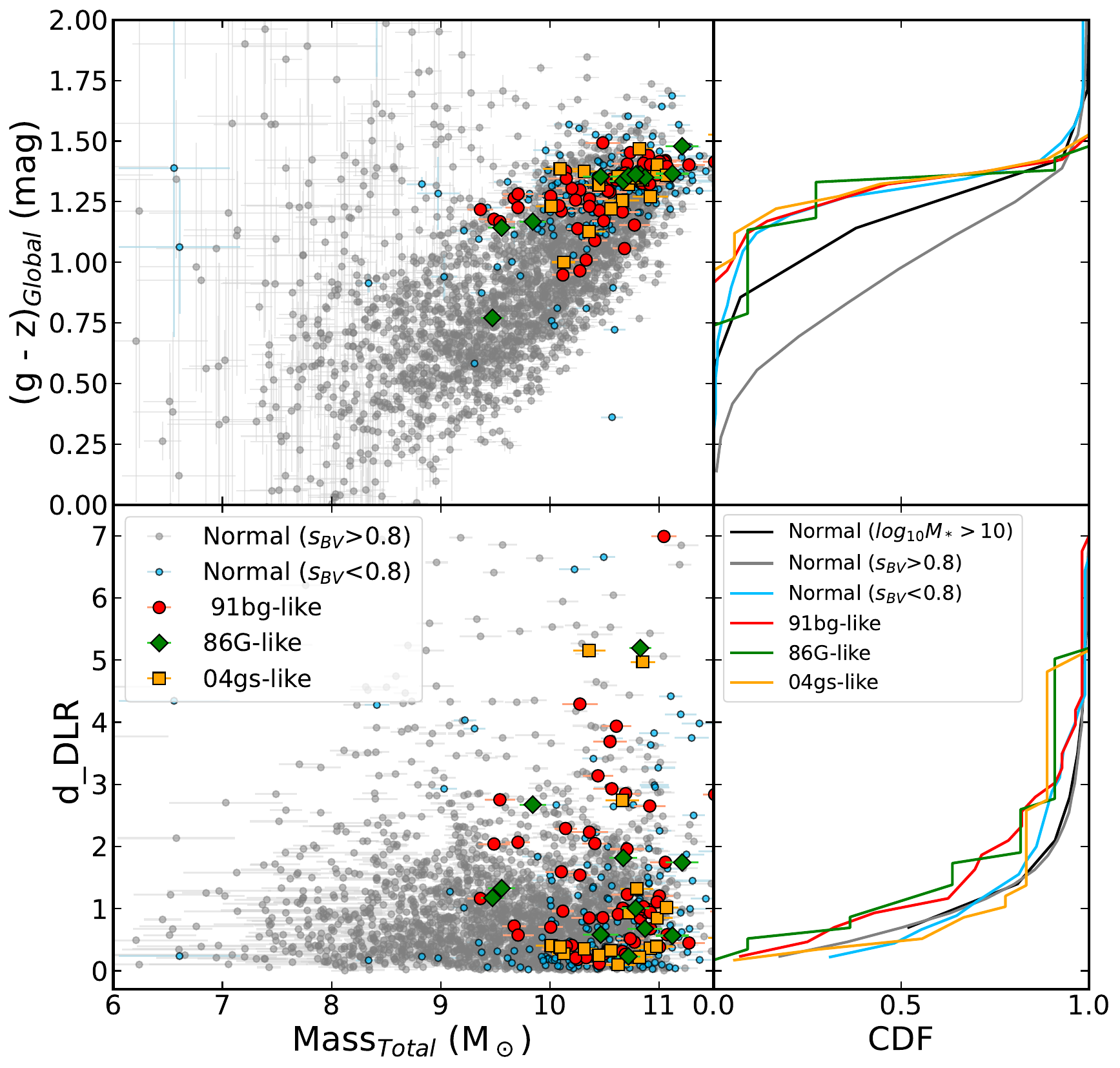}
\caption{Distributions of global host galaxy properties for different SN subtypes. The top-left panel shows the relation between total mass and rest-frame $g-z$ color, while the bottom-left panel illustrates the relation between total mass and the distance-host parameter $d_{DLR}$. The different classifications of SNe shown in the legend include normal SNe Ia from the ZTF DR2 full sample separated by $s_{BV}$ in two subgroups for normal with $s_{BV}$>0.8 in (grey) and normal with $s_{BV}$<0.8 in (skyblue), as well as subluminous subtypes from our sample (91bg-like, 86G-like, and 04gs-like) in (red, green and orange). Cumulative distributions of the host galaxy parameter $d_{DLR}$ are shown in the right side. Also, the distribution of all normal SNe in high-mass galaxies with $log_{10} M_* > 10$ is shown in black.}
\label{fig:global_host}
\end{figure}

\subsubsection{Spectral averages}

We further investigated the differences among subtypes of subluminous and normal SNe by comparing the average spectra of 86G-like, 04gs-like, 91bg-like SNe Ia at maximum light (phase $-$2 < d < +2), alongside the near maximum light spectrum of the normal SN 2011fe, as shown in Fig. \ref{fig:specaverage3}. The most prominent difference among these spectral averages appears in the absorption feature near 4000 \AA~due to \ion{Si}{II} $\lambda$4130 and with increasing contribution of \ion{Ti}{II} (see e.g \citealt{2025A&A...694A...9B}). This feature is clearly distinguishable in the normal and 04gs-like spectra. In contrast, for the 86G-like subtype, it begins to blend with the neighboring \ion{Mg}{II} absorption on the red side, and in 91bg-like SNe, these features merge into a single, broader absorption trough. To emphasize this variation, we zoom in on this region in the right panel of the figure, illustrating the differing strength of the \ion{Ti}{II} absorption feature across the three subtypes. Additional differences between normal and subluminous spectra are clearly observed in the \ion{Si}{II} $\lambda$5972, \ion{O}{I} $\lambda$7774, and \ion{Ca}{II} NIR triplet features. These lines exhibit much shallower absorption in normal SNe Ia spectra compared to their subluminous counterparts.

Additionally, the average continuum of the 04gs-like spectra appears noticeably bluer than that of the 86G-like and 91bg-like subtypes, further highlighting the spectral diversity among subluminous SNe Ia.

\subsection{Host galaxies}\label{subsec:hostgal}

In Fig. \ref{fig:global_host}, we present the relationship between global rest-frame host galaxy color ($g-z$) and $d_{DLR}$, as a function of host galaxy stellar mass. The top-left panel of Fig. \ref{fig:global_host} shows a clear correlation between galaxy color and stellar mass, with galaxies hosting subluminous SNe Ia predominantly located in the upper-right region of the scatter plot (see also \citealt{2025A&A...694A..13B}). These correspond to galaxies with redder colors ($g-z > 1.0$ mag) and high stellar masses ($\log_{10} M_* \gtrsim 10$). In the bottom-left panel of Fig. \ref{fig:global_host}, we observe that more massive galaxies exhibit a broader tail of SNe occurring at larger distances from the galaxy core. Subluminous SNe, particularly the 86G-like and 91bg-like subtypes, appear to follow a similar trend. 

This is further illustrated in the right panels, which show the cumulative distributions of $g-z$ and $d_{DLR}$ for each subluminous SN subtype, as well as for normal SNe Ia. At first glance, the cumulative distributions for 86G- and 91bg-like SNe appear shifted toward larger $d_{DLR}$ values compared to 04gs-like and normal SNe Ia, regardless of whether the latter have $s_{BV}<0.8$, similar to subluminous SNe Ia. This trend persists even when considering only SNe Ia occurring in galaxies with $\log_{10} M_*>10$. To determine whether these visual differences are statistically significant, we performed Kolmogorov-Smirnov (K-S) tests between pairs of distributions. As expected, the 91bg-like distribution shows significant differences, with $p$-values of $<10^{-3}$, 0.002, and 0.001 when compared to normal SNe Ia with $s_{BV}>0.8$, those with $s_{BV}<0.8$, and those in massive galaxies, respectively. These results confirm that 91bg-like SNe Ia tend to occur at larger offsets from the galaxy core. For the 86G-like subtype, the corresponding $p$-values are 0.053, 0.030, and 0.061, which are low but on the threshold of statistical significance in all cases. Meanwhile, the 04gs-like subtype shows no significant differences from normal SNe Ia, with $p$-values of 0.458, 0.261, and 0.406, respectively.  All this is confirmed by the large $p$-value of 0.74 found between 91bg- and 86G-like distributions, and the $p$-value $<0.03$ between 04gs-like and the two 86G- and 91bg-like distributions. In summary, these three subluminous subtypes exhibit a sequence in their spatial distribution relative to normal SNe Ia, with 04gs-like SNe following a distribution similar to normal SNe Ia, while 91bg-like SNe clearly occur at larger distances.

\begin{figure}[t]
\centering
\hspace{-0.7cm}
\includegraphics[width=1.05\columnwidth]{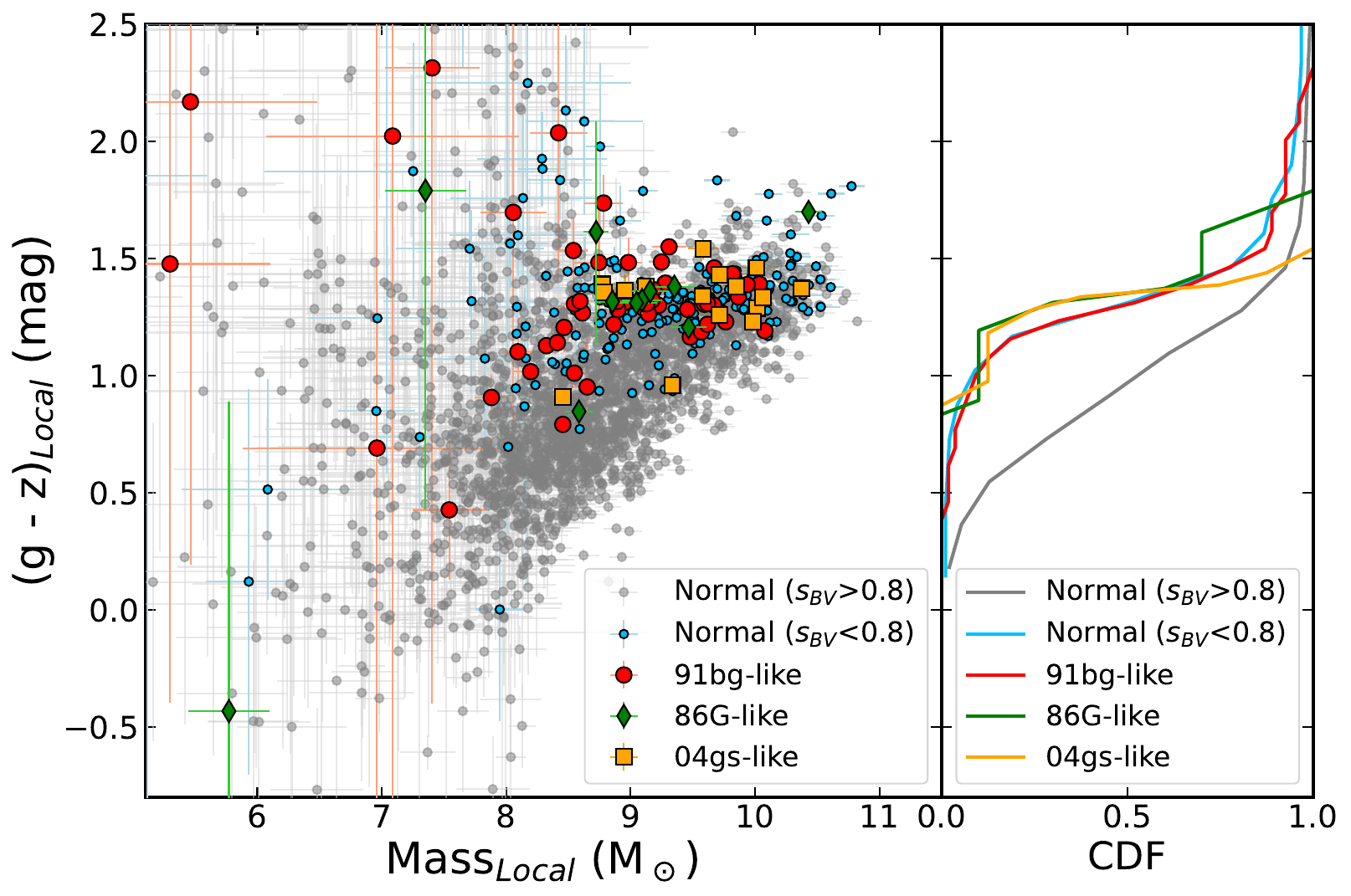}
\caption{Distribution of local host galaxy properties for different SN subtypes shows the correlation between local mass and rest-frame $g-z$ color. The different classifications of SNe shown in the legend include normal SNe Ia from the ZTF DR2 full sample separated by $s_{BV}$ in two subgroups for normal with ($s_{BV}$>0.8) in (grey) and normal with ($s_{BV}$<0.8) in (skyblue), as well as subluminous subtypes from our sample (91bg-like, 86G-like, and 04gs-like) in (red, green and orange), respectively.}
\label{fig:local_host}
\end{figure}

Additionally, Figure \ref{fig:local_host} shows the correlation between local $g - z$ color and stellar mass measured within a 2-kpc diameter aperture centered at the SN location. While both normal and subluminous SNe Ia occur across a similar range of local stellar masses (from 7.5 to 11.5), interestingly, subluminous SNe Ia tend to be systematically located in redder environments for each local mass bin. The observed correlation between redder local color and the occurrence of subluminous SNe Ia suggests that local color 
may effectively trace older stellar populations, which are independently thought to be the progenitors of subluminous SNe Ia.


\section{Discussion}\label{sec:discussion}

\subsection{Intrinsic color effects} 

\begin{figure}[t]
\centering
\hspace*{-0.5cm}
\includegraphics[trim=0.2cm 0.2cm 0.2cm 0.2cm,clip=True,width=\columnwidth]{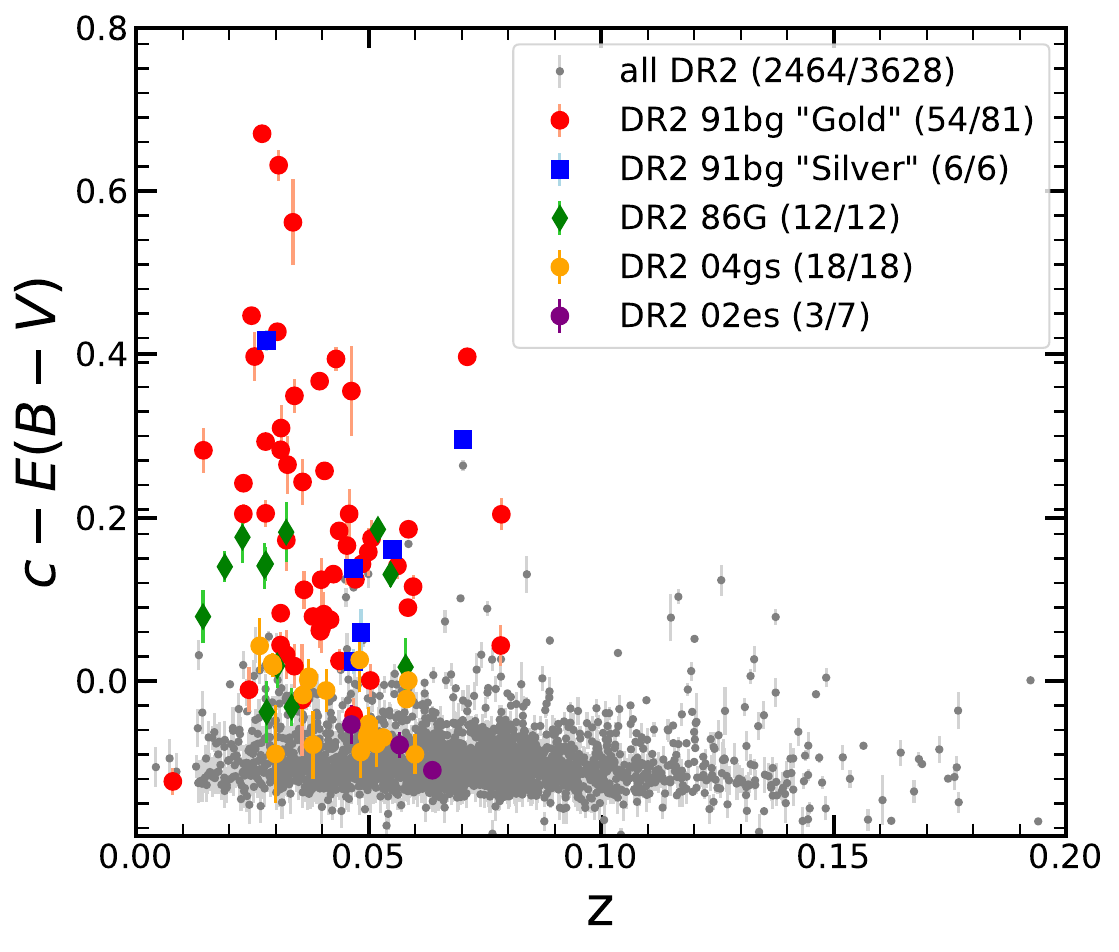}
\caption{A diagram showing the intrinsic color parameter $c$, after subtracting the extrinsic color component $E(B-V)$, as a function of redshift for our subluminous sample and the full ZTF DR2 sample. This isolates intrinsic color effects, revealing that our subluminous SNe Ia tend to be redder, with higher values, compared to normal SNe Ia. We note that the number of SNe included is reduced for both normal and subluminous subtypes, as this analysis requires reliable fits from both SALT2 and SNooPy.}
\label{fig:cebvdiff}
\end{figure}

In Fig. \ref{fig:cebvdiff}, we show the difference between the SALT2 color parameter, $c$, which reflects both intrinsic and extrinsic contributions, and the SNooPy color excess, $E(B-V)$, which traces only extrinsic reddening. While $c$ is not a strict linear sum of intrinsic and extrinsic color components, several recent works (e.g. \citealt{2021ApJ...909...26B}; \citealt{2022MNRAS.515.4587W}; \citealt{2023ApJ...945...84P}) show that, to first order, the observed color can be parameterized as $c_{obs} \approx c_{int}+E(B-V)$. Thus, subtracting the SNooPy reddening provides a color term effectively related to the intrinsic color component, though we emphasize that simulations would be necessary to quantify possible biases from differences between $E(B-V)_{SALT2}$ and $E(B-V)_{SNooPy}$. We then plot this intrinsic-related term as a function of redshift for both our subluminous sample and the full ZTF DR2 sample. Our results confirm that SNe Ia in the subluminous sample tend to be intrinsically redder than normal SNe Ia. These redder colors are primarily attributed to their lower temperatures, resulting in cooler photospheres and stronger \ion{Ti}{II} absorption features, suggesting that subluminous SNe Ia represent the low end of a temperature sequence.

We calculated the weighted averages of the y-axis $(c - E(B-V))$ for different subtypes of SNe Ia up to a redshift of $z \leq 0.06$, which corresponds to the completeness limit for normal SNe Ia in our sample. The results show that 91bg-like SNe, with a weighted average of 0.43$\pm$0.03 mag, are significantly redder than all other subtypes \citep{1992AJ....104.1543F}. This differs notably from normal SNe, which have an average of $-$0.03$\pm$0.04 mag, and from 04gs-like SNe, with an average of $-$0.02$\pm$0.04 mag — indicating that 91bg-like SNe form a clearly distinct population. 86G-like SNe also appear redder, with an average value of 0.13$\pm$0.07 mag, but the difference relative to normal and 04gs-like SNe corresponds to only $\sim$ 1.9$\sigma$, which does not provide strong evidence for a clear separation. In contrast, the color difference between 04gs-like and normal SNe Ia is minimal (< 0.01 mag,  $\ll$1$\sigma$), suggesting that this similarity in intrinsic colour may hint at, though does not by itself confirm, a possible connection between the progenitor systems of these two subtypes. Overall, the analysis highlights that 04gs-like and 86G-like SNe remain much closer to normal SNe Ia, whereas only the 91bg-like SNe Ia stand out as statistically distinct, with >9.6$\sigma$ difference compared to normal/04gs-like and 4$\sigma$ compared to 86G-like. This is broadly consistent with the diversity among SNe discussed in \citet{2025A&A...694A...9B}, which suggested from spectral point of view a continuum extending from bright to faint normal SNe Ia, including the transitional and 91bg-like SNe Ia, suggesting that variations within a single explosion model could potentially account for the observed diversity. However, our results show a clear lack of continuity in the photometric view, as only the 91bg-like SNe Ia appear significantly redder and distinct once intrinsic and extrinsic color contributions are disentangled, while the other subtypes form a continuum.

This trend is robust against possible external biases. It is independent of 91bg-like SNe Ia occurrence in redder environments compared to normal SNe Ia, as discussed in Section \ref{subsec:hostgal}, since SN observables are derived directly from the light-curve fit. Also their classification in subtypes was performed based on spectral features, also independent of SN light-curve parameters, ensuring that the observed photometric separation is intrinsic.

\begin{figure*}[!t]
\centering
\includegraphics[width=0.7\textwidth]{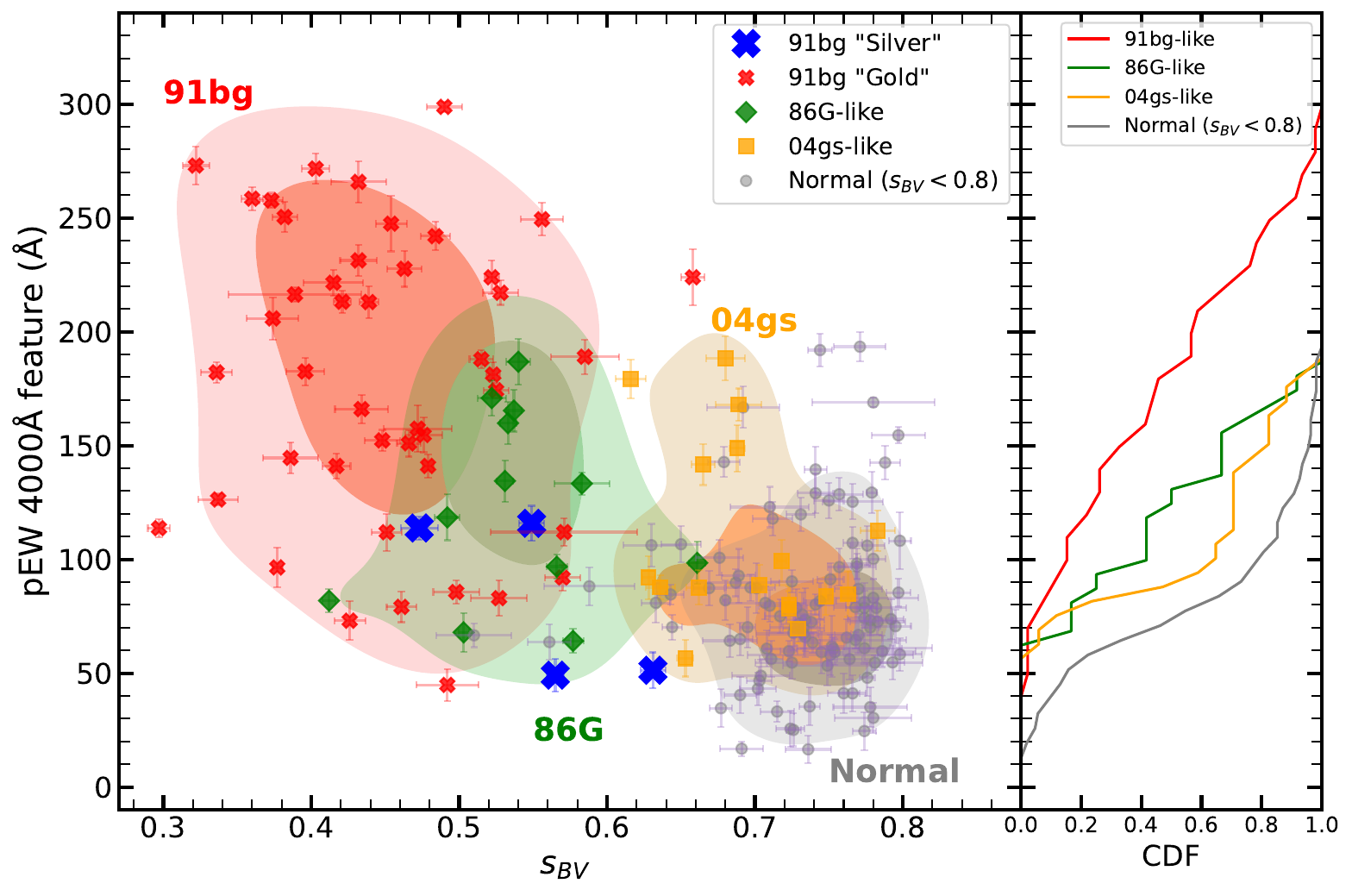}
\caption{The pEW diagram of \ion{Ti}{II} $\lambda4000$ plotted against the color stretch parameter $s_{BV}$ showing the three subtypes ‘91bg-like’ (red), ‘86G-like’ (green), and ‘04gs-like’ (orange) from our subluminous sample, along with normal SNe Ia with $s_{BV}$ < 0.8 (grey) as distinguished different SNIa populations based on their spectral characteristics and color evolution. The shaded regions around each subtype indicate the confidence levels ($1\sigma$ and $2\sigma$ ) of their distributions in the pEW–$s_{BV}$ plane. Additionally, the 4 ‘91bg-like’ SNe Ia labeled as "Silver" candidates in (blue)  are plotted and re-classified with respect to their distance to maximum of the KDE distributions of the three SN Ia populations (91bg-like, 86G-like and 04gs-like). The cumulative distributions for all different SN Ia populations are shown in the right panel.}
\label{fig:tipew}
\end{figure*}

\subsection{pEW of the spectral feature \ion{Ti}{II} $\lambda$4000}\label{subsec:pewTiII}

In this section, we discuss the distribution of the $pEW$ of the spectral feature \ion{Ti}{II} $\lambda4000$, highlighting its importance in distinguishing between different SN populations (91bg-like, 86G-like, and 04gs-like), as previously shown in Fig. \ref{fig:specaverage3}. We used {\sc Spextractor} to estimate the $pEW$ of the whole \ion{Ti}{ii}, \ion{Si}{ii} and \ion{Mg}{ii} feature, treating them as a single broad absorption, defining the continuum on the blueward side of the \ion{Ti}{ii} feature and on the redward side of the \ion{Mg}{ii} feature. 

Distinct SN Ia populations can be clearly visualized in the right panel of Fig. \ref{fig:tipew}, where the cumulative distributions of the $pEW$ of this broad feature are presented for the subluminous subtypes and the normal SNe Ia with $s_{BV}<0.8$, typical of the subluminous subtypes. In this case, the K-S $p$-values of the three subluminous subtypes 91bg-, 86G- and 04gs-like are low enough when compared to the normal SNe Ia with $s_{BV}$ ($<10^{-3}$, 0.003, 0.006, respectively) to conclude that this feature distinctively identifies these subluminous SN Ia subtypes. Similarly, the K-S $p$-value between the 86G- and 04gs-like distributions is 0.313, pointing to similar underlying distributions, although the cumulative distribution of 86G-likes shown in the right side of Fig. \ref{fig:tipew} is shifted to higher values. Finally, when comparing the 91bg-like distribution to both other subtypes we obtain $p$-values of 0.016 for 86G- and 0.002 with 04gs-like. The differences between these distributions underscore the significant role of the $\lambda$4000 \AA~spectral feature in distinguishing between these subtypes and provide strong evidence that ‘86G-like’ and ‘04gs-like’ SNe Ia are distinct from ‘91bg-like’ SNe Ia. Furthermore, the left panel of Fig.\ref{fig:tipew} presents the $pEW$ of this combined $\lambda4000$ \AA~feature as a function of the color-stretch $s_{BV}$ parameter. This diagram reveals variations across different SN subtypes. Specifically, 91bg-like SNe Ia Gold exhibit a strong absorption and associated with low $s_{BV}$ values, characteristic of cooler and redder SNe Ia. In contrast, normal SNe Ia typically display a weaker absorption and higher $s_{BV}$ values. 86G-like and 04gs-like SNe Ia fall between these two extremes, showing moderate \ion{Ti}{II} absorption. 

To further investigate the effect of overlapping values and the potential separation between the three subtypes and low $s_{BV}$ normal SNe Ia, we analyzed the density of the points in the diagram using a kernel density estimate (KDE) and generate bi-variate distributions that represents the data through continuous probability density curves. Additionally, we examined the four ‘91bg-like’ SNe Ia labeled as {\it Silver} candidates to refine their sub-classification in relation to the identified SN Ia populations in the pEW $\lambda$4000 \AA~diagram. From the KDE distributions we estimated the probability density function (PDF) values at the positions of the four 91bg-like Silver SNe Ia candidates. Subsequently, we calculated the differences between these density estimates and their respective KDE maxima, expressing these differences in units of standard deviations. This approach allowed us to reclassify the four ‘Silver’ SNe Ia candidates based on their density deviations relative to the three SN populations. Table \ref{tab:KDEparams} presents these KDE values as sigma measurements, representing the deviation of each 'Silver' SN candidate from the peak density distribution of each SN population. Finally, we used these sigma values as a criterion to determine the most likely SN population to which each candidate belongs, based on how closely they align with a specific SN population. The significance of the sigma values can be interpreted as follows: the lowest value for each 'Silver' SN candidate indicates the closest match to a specific SN population. Based on this criterion, we reclassified ZTF19acavtco as a ‘91bg-like’ SN Ia, ZTF18abbikrz and ZTF20aagloch as ‘86G-like’ SNe Ia, and ZTF19adcdgca as a ‘04gs-like’ SN Ia.

\begin{figure*}[t]
\centering
\hspace*{-0.4cm}
\includegraphics[width=\textwidth]{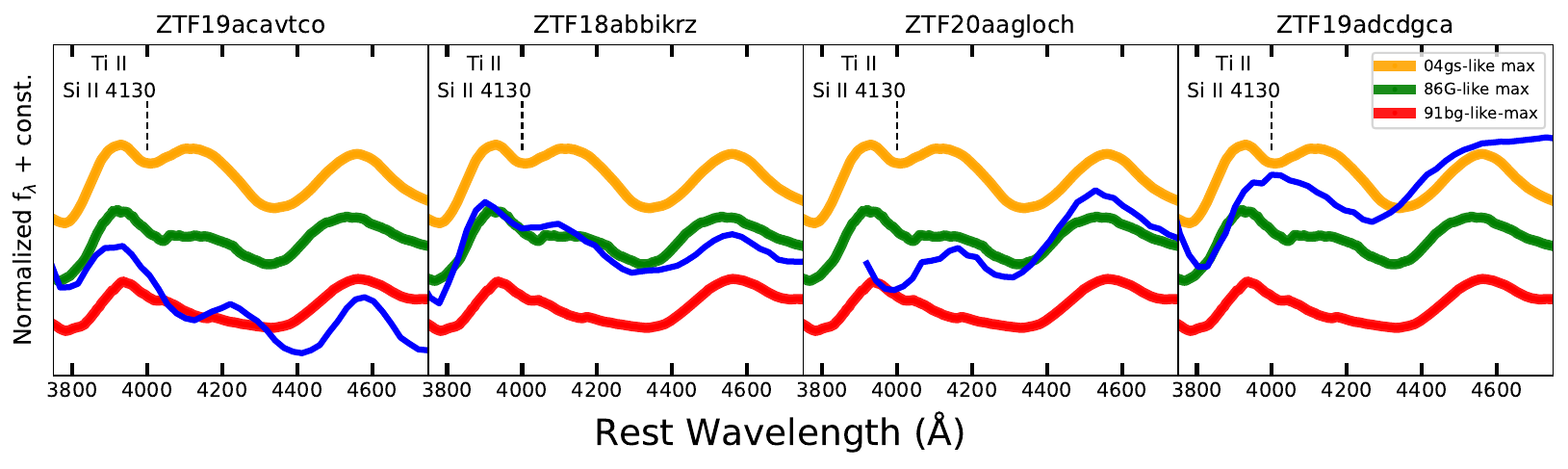}
\caption{A comparison of the 4 SNe Ia (ZTF19acavtco, ZTF18abbikrz, ZTF20aagloch and ZTF19adcdgca) classified previously as ‘91bg-like’ "Silver" candidates, focusing on their Ti II absorption, plotted together with the corresponding spectral average of each sub-type, which shows how well each of these re-classified SNe, based the calculated KDE results presented in \ref{subsec:pewTiII}, matches with its new SN Ia subtype (91bg-like, 86G-like and 04gs-like).} 
\label{fig:91bgsilver_pewti}
\end{figure*}

Finally, as shown in Fig. \ref{fig:91bgsilver_pewti}, we compared the region of the spectra around the $\lambda4000$ \AA~absorption spectral feature of these four reclassified SNe Ia with the spectral averages of the three subluminous subtypes. This comparison aims to verify how closely our reclassification aligns with the spectral reference averages concerning the \ion{Ti}{II} feature, ensuring the accuracy of our results. From the comparison we can confirm that the 4 re-classified SNe Ia match very well with its corresponding spectral average of each SN sub-subtype, confirming our re-classification based the calculated KDE results.

However, upon visual examination by eye of Fig. \ref{fig:91bgsilver_pewti} we note that, apart from ZTF18abbikrz, the reclassifications of the other three SNe are less robust, as they exhibit similarities with alternative subtypes. In particular, ZTF19acavtco appears consistent with both the ‘91bg-like’ and ‘86g-like’ spectral averages, ZTF20aagloch shows affinities with both the ‘86g-like’ and ‘04gs-like’, and ZTF19adcdgca lies intermediate between the ‘04gs-like’ and ‘91bg-like’.
\begin{table}[t]
\caption{KDE values of standard deviation for the 4 ‘91bg-like’ SNe Ia "Silver" measured with respect to the SN Ia population KDE maxima (See, Fig. \ref{fig:tipew}).}
\label{tab:KDEparams}
\centering
\setlength{\tabcolsep}{5pt}
\resizebox{1.0\columnwidth}{!}{
\begin{tabular}{l|ccccc}
\hline\hline
\begin{tabular}[c]{@{}l@{}}ZTF Name\end{tabular} & \multicolumn{1}{c}{\begin{tabular}[c]{@{}c@{}}$s_{BV}$\end{tabular}} & \multicolumn{1}{c}{\begin{tabular}[c]{@{}c@{}}pEW (\ion{Ti}{II} $\lambda$4000)\\ ($\textup{\r{A}}$)\end{tabular}} & \multicolumn{1}{c}{\begin{tabular}[c]{@{}c@{}}$\sigma_{91bg}$\end{tabular}}& \multicolumn{1}{c}{\begin{tabular}[c]{@{}c@{}}$\sigma_{86G}$\end{tabular}}&\multicolumn{1}{c}{\begin{tabular}[c]{@{}c@{}}$\sigma_{04gs}$\end{tabular}} 
\\
\hline
ZTF19acavtco & 0.473 & 113.75 & 0.157$\sigma$   & 0.447$\sigma$  & 1.335$\sigma$ \\
ZTF18abbikrz & 0.549 & 115.81 & 0.485$\sigma$   & 0.102$\sigma$  & 1.160$\sigma$\\
ZTF20aagloch & 0.565 & 49.11  & 1.014$\sigma$   & 0.470$\sigma$  & 1.107$\sigma$ \\
ZTF19adcdgca & 0.631 & 51.26  & 1.516$\sigma$   & 0.710$\sigma$  & 0.587$\sigma$\\
\hline
\end{tabular}
}
\end{table}


\section{Summary and conclusions}\label{sec:conclusion}

The distribution of light-curve parameters from SALT2 $x_1$ vs. $c$ and from SNooPy $s_{BV}$ vs. $E(B-V)$ revealed key differences between subluminous and normal SNe Ia, where subluminous SNe Ia generally exhibited lower stretch $x_1$ values and higher color $c$ values, indicating faster-evolving, redder light curves compared to normal SN Ia (\citealt{1993ApJ...413L.105P}; \citealt{2014ApJ...789...32B}). Similarly, the color stretch $s_{BV}$ further differentiate these subluminous SNe Ia from normal SNe Ia, but the $E(B-V)$ distributions appeared more similar. 02es-like SNe Ia were always found at widths ($x_1$ and $s_{BV}$) compatible to normal SNe Ia, but at high $c$ and $E(B-V)$ values. Both light-curve width vs. color/extinction-corrected luminosity relations show a very well-defined linear behavior of normal SNe Ia, while subluminous SNe Ia appear below the relation for normal SNe Ia. SALT2 accounts for that difference in brightness within the color correction, making all subluminous SNe Ia following the relation of normal SNe Ia to low $x_1$ values. Since the SALT2-$c$ parameter accounts for both intrinsic and extrinsic color differences, SALT2 may be artificially overcorrecting subluminous SNe Ia assuming they are just normal SNe Ia with narrower light-curves affected by high extinction. Given that the color–luminosity relation of fast-declining SNe Ia only coincidentally resembles dust reddening, this approach risks introducing hidden systematics into cosmological distance measurements, including determinations of $H_0$.

In this work, we perform a detailed photometric and spectroscopic analysis of 124 subluminous SNe~Ia from ZTF DR2, the largest sample of spectroscopically classified subluminous type Ia supernova observed with a single instrument. This covers all subluminous SN Ia subtypes, including the 91bg-like subtype, 86G-like, 04gs-like, and 02es-like. On the photometric side, we fit the light curves using \textsc{SNooPy} to extract key parameters such as ($s_{BV}$, $E(B-V)$, $t_{max}$ and $m_{max}$), and combine these results with \textsc{SALT2} parameters provided in the published ZTF DR2 tables. For \textsc{SNooPy}, we complement ZTF photometry with publicly available ATLAS light curves, significantly improving estimates of the time of maximum light, in particular for objects lacking early-time pre-maximum ZTF observations. On the spectroscopic side, we use \textsc{Spextractor} to measure expansion velocities and pseudo-equivalent widths ($pEW$) of spectral features in 194 available spectra for the 124 SNe~Ia. We focus particularly on the \ion{Si}{II}~$\lambda$6355 and \ion{Si}{II}~$\lambda$5972 lines and define a broad absorption feature around 4300~\AA\ dominated by \ion{Ti}{II}, \ion{Si}{II}, and \ion{Mg}{II}, which we use to differentiate subluminous subtypes. We determine the spectral phases for the full sample and construct binned average spectra to compare the subtypes at maximum light. Additionally, we obtain from the ZTF DR2 global and local 2~kpc circular aperture photometry of the host galaxies, along with galactocentric distances in units of directional light radius ($d_{DLR}$), to study differences in host galaxy stellar mass, $g - z$ color, and explosion site among subluminous and normal SNe~Ia.

Distributions in light-curve parameters such as $x1$ vs. $c$ and $s_{BV}$ vs. $E(B-V)$ show that subluminous SNe~Ia generally have lower stretch and redder colors than normal SNe~Ia, with the exception of 02es-like events, which have light-curve widths comparable to normal SNe~Ia but higher $c$ and $E(B-V)$ values. Both the width–color and extinction-corrected luminosity relations show a tight linear correlation for normal SNe~Ia. Sub-luminous events fall below this relation, but \textsc{SALT2} compensates for their fainter luminosities by increasing the $c$ parameter, effectively aligning them with the standard luminosity-width relation through artificially low $x1$ values. Since \textsc{SALT2}-$c$ accounts for both intrinsic and extrinsic color, this may lead to an overcorrection, treating subluminous SNe~Ia as normal events with narrower light curves and high extinction.
To explore intrinsic color differences, we subtract the extrinsic color estimated from \textsc{SNooPy} $E(B-V)$ from the color parameter \textsc{SALT2}-$c$, which incorporate both extrinsic and intrinsic effects. While normal SNe~Ia cluster around $c-E(B-V)\sim-0.1$~mag, subluminous SNe~Ia span a broader range from $-0.1$ to $+0.7$~mag.

Most (90\%) of the subluminous SNe~Ia with $pEW$ measurements of \ion{Si}{II} taken within $-5 < t < +5$ days fall within the \textit{Cool} region of the \citet{2006PASP..118..560B} classification diagram, as expected. The exceptions, four 02es-like and five 91bg-like SNe, exhibit lower \ion{Si}{II}~$\lambda$5972 values, which we attribute to host galaxy contamination or suboptimal observing conditions and reductions. The strength of the blended absorption near 4300~\AA\ (dominated by \ion{Ti}{II}) serves as a reliable discriminator of subluminous spectral diversity, following a sequence from shallower 04gs-like to deeper 91bg-like. The $pEW_{\lambda4300}$ versus $s_{BV}$ diagram provides a powerful tool for distinguishing these subtypes among themselves and from normal SNe~Ia with narrow light curves ($s_{BV} < 0.8$). As a proof, by using kernel density estimation (KDE), we classify the four Silver 91bg-like SNe Ia with spectra available by computing distances to the KDE peaks of each subtype. We find that all subluminous SNe~Ia preferentially occur in more massive, redder host galaxies and explode in the reddest local environments for a given mass bin, consistent with progenitors originating in older stellar populations. Moreover, 91bg-like and 86G-like events are found at significantly larger galactocentric distances.

\section*{Data availability}
The full version of Tables \ref{tab:LC_parameters} and \ref{tab:vel-pew} are only available in electronic form at the CDS via anonymous ftp to \href{http://cdsarc.u-strasbg.fr}{cdsarc.u-strasbg.fr (130.79.128.5)} or via \href{http://cdsweb.u-strasbg.fr/cgi-bin/qcat?J/A+A/}{http://cdsweb.u-strasbg.fr/cgi-bin/qcat?J/A+A/}.

\begin{acknowledgements}
This work has been carried out within the framework of
the doctoral program in Physics of the Universitat Autònoma de Barcelona. The SNICE research group acknowledges financial support from AGAUR, CSIC, MCIN, and AEI (10.13039/501100011033) under projects PID2023-151307NB-I00, PIE 20215AT016, CEX2020-001058-M, ILINK23001, COOPB2304, and 2021-SGR-01270.
U.B. and T.E.M.B. are funded by Horizon Europe ERC grant no. 101125877.
This work has been supported by the research project grant “Understanding the Dynamic Universe” funded by the Knut and Alice Wallenberg Foundation under Dnr KAW 2018.0067 and the {\em Vetenskapsr\aa det}, the Swedish Research Council, project 2020-03444.
Y.-L.K. was supported by the Lee Wonchul Fellowship, funded through the BK21 Fostering Outstanding Universities for Research (FOUR) Program (grant No. 4120200513819) and the National Research Foundation of Korea to the Center for Galaxy Evolution Research (RS-2022-NR070872, RS-2022-NR070525).
This project has received funding from the European Research Council (ERC) under the European Union's Horizon 2020 research and innovation program (grant agreement n 759194 - USNAC).
Based on observations obtained with the Samuel Oschin Telescope 48-inch and the 60-inch Telescope at the Palomar Observatory as part of the Zwicky Transient Facility project. ZTF is supported by the National Science Foundation under Grants No. AST-1440341 and AST-2034437 and a collaboration including current partners Caltech, IPAC, the Weizmann Institute of Science, the Oskar Klein Center at Stockholm University, the University of Maryland, Deutsches Elektronen-Synchrotron and Humboldt University, the TANGO Consortium of Taiwan, the University of Wisconsin at Milwaukee, Trinity College Dublin, Lawrence Livermore National Laboratories, IN2P3, University of Warwick, Ruhr University Bochum, Northwestern University and former partners the University of Washington, Los Alamos National Laboratories, and Lawrence Berkeley National Laboratories. Operations are conducted by COO, IPAC, and UW. SED Machine is based upon work supported by the National Science Foundation under Grant No. 1106171. The ZTF forced-photometry service was funded under the Heising-Simons Foundation grant \#12540303 (PI: Graham). This work was supported by the GROWTH project funded by the National Science Foundation under Grant No 1545949 \citep{Kasliwal2019pasp}. Fritz \citep{vanderWalt2019, Coughlin2020apjs} is used in this work. The Gordon and Betty Moore Foundation, through both the Data-Driven Investigator Program and a dedicated grant, provided critical funding for SkyPortal.
\end{acknowledgements}

\bibliographystyle{aa}
\bibliography{ztf} 

\begin{appendix}
\onecolumn
\section{Tables}

\begin{table}[h]
\caption{SN Ia light-curve parameters.}
\centering
\setlength{\tabcolsep}{4pt}
\resizebox{0.95\columnwidth}{!}{
\begin{tabular}{ll|cccc|ccccc}
\hline\hline
&\multicolumn{1}{c|}{} 
&\multicolumn{4}{c|}{SALT2}
&\multicolumn{5}{c}{SNooPy}\\

\hline\hline
\begin{tabular}[c]{@{}l@{}}ZTF Name\end{tabular} &
\begin{tabular}[c]{@{}l@{}}IAU Name\end{tabular} &
\multicolumn{1}{c}{\begin{tabular}[c]{@{}c@{}} $t^{B}_{\rm max}$ \\ (MJD) \end{tabular}} &
\multicolumn{1}{c}{\begin{tabular}[c]{@{}c@{}} $B_{\rm max}$ \\ (mag) \end{tabular}} &
\multicolumn{1}{c}{\begin{tabular}[c]{@{}c@{}} $x_1$ \\  \end{tabular}}&
\multicolumn{1}{c|}{\begin{tabular}[c]{@{}c@{}} $c$  \\ (mag) \end{tabular}} &
\multicolumn{1}{c}{\begin{tabular}[c]{@{}c@{}} $t^{G}_{\rm max}$ \\ (MJD) \end{tabular}} &
\multicolumn{1}{c}{\begin{tabular}[c]{@{}c@{}} $G_{\rm max}$ \\ (mag) \end{tabular}} &
\multicolumn{1}{c}{\begin{tabular}[c]{@{}c@{}} $s_{BV}$ \\  \end{tabular}} &
\multicolumn{1}{c}{\begin{tabular}[c]{@{}c@{}} $E(B-V)$ \\ (mag) \end{tabular}} &
\multicolumn{1}{c}{\begin{tabular}[c]{@{}c@{}}Source$^{(1)}$\end{tabular}} \\
\hline
ZTF18aahfgyz   &   2018ast   &  58215.87$\pm$0.295 &  17.24$\pm$0.277   &  	-2.08$\pm$0.087    & 	0.537$\pm$0.038 &	58215.28$\pm$0.26   &  17.64$\pm$0.400  &  	0.322$\pm$0.018    &  	0.237$\pm$0.032 & ZTF\\
ZTF18aahjaxz   &   2018avg   &  58226.16$\pm$0.151 &  18.97$\pm$0.228   &  	-3.04$\pm$0.159    & 	0.340$\pm$0.037 &	58224.12$\pm$0.33   &  19.62$\pm$0.427  &  	0.476$\pm$0.025    &   -0.014$\pm$0.065 & ZTF\\
ZTF18aaimxdx   &   2018bay   &  58217.23$\pm$0.117 &  18.86$\pm$0.269   &  	-2.58$\pm$0.100    & 	0.460$\pm$0.038 &	58217.61$\pm$0.13   &  18.96$\pm$0.620  &  	0.454$\pm$0.021    &  	0.066$\pm$0.046 & ZTF\\
ZTF18aajtlbf   &   2018bbz   &  58228.08$\pm$0.096 &  17.74$\pm$0.199   &  	-3.24$\pm$0.088    & 	0.337$\pm$0.032 &	58227.07$\pm$0.12   &  17.53$\pm$0.413  &  	0.531$\pm$0.011    &  	0.194$\pm$0.022 & ZTF\\
ZTF18aaoxrup   &   2020hdw   &  58955.30$\pm$0.121 &  17.47$\pm$0.191   &  	-1.79$\pm$0.060    & 	0.148$\pm$0.029 &	58956.92$\pm$0.12   &  17.44$\pm$0.000  &  	0.653$\pm$0.011    &  	0.160$\pm$0.016 & ZTF\\
ZTF18aarcypa   &   2018bil   &  58252.46$\pm$0.339 &  18.23$\pm$0.495   &  	-3.04$\pm$0.290    & 	0.151$\pm$0.071 &	58252.44$\pm$0.82   &  16.94$\pm$0.326  &  	0.571$\pm$0.099    &  	0.105$\pm$0.123 & ZTF\\
ZTF18aaroihe   &   2018bio   &  58259.26$\pm$0.079 &  17.67$\pm$0.189   &  	-1.96$\pm$0.072    &   -0.045$\pm$0.029 &	58259.43$\pm$0.06   &  17.63$\pm$0.647  &  	0.763$\pm$0.008    &  	0.042$\pm$0.013 & ZTF\\
ZTF18aasprui   &   2018euz   &  58347.54$\pm$0.093 &  17.41$\pm$0.176   &  	-2.01$\pm$0.048    &   -0.005$\pm$0.027 &	58347.85$\pm$0.07   &  17.37$\pm$0.056  &  	0.748$\pm$0.004    &  	0.073$\pm$0.007 & ZTF\\
ZTF18aayiahw   &   2018cts   &  58289.47$\pm$0.214 &  19.38$\pm$0.379   &  	-0.29$\pm$0.423    & 	0.396$\pm$0.050 &	58289.67$\pm$0.35   &  19.11$\pm$0.087  &  	1.180$\pm$0.061    &  	0.506$\pm$0.048 & ZTF\\
ZTF18abbikrz   &   2020pwn   &  59058.05$\pm$0.097 &  17.65$\pm$0.207   &  	-2.83$\pm$0.149    & 	0.009$\pm$0.032 &	59058.24$\pm$0.10   &  17.61$\pm$0.200  &  	0.549$\pm$0.017    &   -0.128$\pm$0.033 & ZTF\\
ZTF18abdmgab   &   2018lph   &  58311.52$\pm$0.086 &  19.38$\pm$0.179   &  	-2.41$\pm$0.122    & 	0.087$\pm$0.029 &	58311.41$\pm$0.12   &  19.26$\pm$0.398  &  	0.658$\pm$0.016    &  	0.044$\pm$0.017 & ZTF\\
ZTF18abixkdo   &   2018eak   &  58328.84$\pm$0.155 &  17.03$\pm$0.212   &  	-2.12$\pm$0.136    & 	0.043$\pm$0.034 &	58329.11$\pm$0.09   &  17.12$\pm$0.050  &  	0.661$\pm$0.011    &  	0.025$\pm$0.021 & ZTF\\
ZTF18abltdfj   &   2018eyi   &  58340.00$\pm$0.022 &  19.31$\pm$0.245   &  	-1.88$\pm$0.107    & 	0.712$\pm$0.035 &	58339.26$\pm$0.21   &  18.92$\pm$0.000  &  	0.360$\pm$0.015    &  	0.285$\pm$0.036 & ZTF\\
ZTF18ablwtkf   &   2018ezz   &  58347.29$\pm$0.120 &  17.66$\pm$0.188   &  	-2.34$\pm$0.052    & 	0.119$\pm$0.029 &	58347.95$\pm$0.08   &  17.68$\pm$0.084  &  	0.503$\pm$0.007    &   -0.063$\pm$0.011 & ZTF\\
ZTF18abmjcbs   &   $\cdots$  &  58349.13$\pm$0.171 &  18.43$\pm$0.253   &  	-2.93$\pm$0.165    & 	0.211$\pm$0.037 &	58347.90$\pm$0.19   &  19.70$\pm$1.443  &  	0.522$\pm$0.019    &  	0.026$\pm$0.032 & ZTF\\
$\vdots$   &   $\vdots$   &  $\vdots$ &  $\vdots$   &  	$\vdots$    & 	$\vdots$ &	$\vdots$   &  $\vdots$  &  	$\vdots$    &  	$\vdots$ & $\vdots$ \\

\hline

\end{tabular}}
\label{tab:LC_parameters}
\end{table}


\begin{table}[h]
\caption{Spectroscopic properties of all 124 SNe in the initial sample, including \ion{Si}{II} $\lambda$6355 velocity and pseudo-equivalent widths of key features (\ion{Si}{II} $\lambda$6355, \ion{Si}{II} $\lambda$5972, \ion{Mg}{II} $\lambda$4000, \ion{O}{I} $\lambda$7774, and \ion{Ca}{II} NIR), together with the spectral phase and spectroscopic classification of each SN.}
\centering
\setlength{\tabcolsep}{3pt}
\label{tab:vel-pew}
\resizebox{1.0\textwidth}{!}{
\begin{tabular}{ll|cccccccc}
\\
\hline \hline
\\
\begin{tabular}[c]{@{}l@{}}ZTF Name\end{tabular} &
\multicolumn{1}{c}{\begin{tabular}[c]{@{}c@{}}IAU Name\end{tabular}} &
\multicolumn{1}{c}{\begin{tabular}[c]{@{}c@{}}Subtype\end{tabular}} &
\multicolumn{1}{c}{\begin{tabular}[c]{@{}c@{}}Phase \\ (days)\end{tabular}}  &
\multicolumn{1}{c}{\begin{tabular}[c]{@{}c@{}}v (\ion{Si}{II} $\lambda$6355)\\ ($10^{3}$ km s$^{-1}$) \end{tabular}} &
\multicolumn{1}{c}{\begin{tabular}[c]{@{}c@{}}pEW (\ion{Si}{II} $\lambda$6355)\\ ($\textup{\r{A}}$)\end{tabular}} &
\multicolumn{1}{c}{\begin{tabular}[c]{@{}c@{}}pEW (\ion{Si}{II} $\lambda$5972)\\ ($\textup{\r{A}}$)\end{tabular}}&
\multicolumn{1}{c}{\begin{tabular}[c]{@{}c@{}}pEW (\ion{Mg}{II} $\lambda$4000)\\ ($\textup{\r{A}}$)\end{tabular}}&
\multicolumn{1}{c}{\begin{tabular}[c]{@{}c@{}}pEW (\ion{O}{I} $\lambda$7774)\\ ($\textup{\r{A}}$)\end{tabular}}&
\multicolumn{1}{c}{\begin{tabular}[c]{@{}c@{}}pEW (\ion{Ca}{II} NIR)\\ ($\textup{\r{A}}$)\end{tabular}}
\\
\hline \hline

ZTF18aahfgyz &2018ast    &91bg-like ‘Gold’ &  2.102   &   9.539 $\pm$ 0.443 & 81.183 $\pm$ 3.824 & 46.15 $\pm$ 3.612 & 276.189 $\pm$ 4.284 & 100.153 $\pm$ 2.74 & 273.212 $\pm$ 2.809        \\
ZTF18aahjaxz &2018avg    &91bg-like ‘Gold’ & -0.157   &   9.965 $\pm$ 0.436 & 80.674 $\pm$ 2.976 & 36.89 $\pm$ 3.246 & 131.721 $\pm$ 3.93 & 96.604 $\pm$ 1.557 & 371.657 $\pm$ 1.7\\
ZTF18aaimxdx &2018bay    &91bg-like ‘Gold’ &  1.693   &   9.022 $\pm$ 0.375 & 107.956 $\pm$ 3.37 & 41.864 $\pm$ 3.136 & 255.711 $\pm$ 6.676 & 132.354 $\pm$ 2.018 & 300.368 $\pm$ 2.391\\
ZTF18aarcypa &2018bil    &91bg-like ‘Gold’ &  1.490   &   9.548 $\pm$ 0.7 & 75.937 $\pm$ 3.755 & 22.192 $\pm$ 3.777 & $\cdots$ & 43.373 $\pm$ 2.018 & 74.142 $\pm$ 1.676\\
ZTF18abdmgab &2018lph    &91bg-like ‘Gold’ &  0.437   &   10.279 $\pm$ 0.331 & 135.307 $\pm$ 2.964 & 46.826 $\pm$ 2.818 & 242.061 $\pm$ 6.778 & 133.78 $\pm$ 1.311 & 379.614 $\pm$ 1.857\\
ZTF18abltdfj &2018eyi    &91bg-like ‘Gold’ &  0.970   &    8.804 $\pm$ 0.69 & 92.554 $\pm$ 4.056 & 32.483 $\pm$ 4.112 & 270.29 $\pm$ 2.643 & 121.536 $\pm$ 2.625 & 294.415 $\pm$ 2.559\\
ZTF18abnzocn &2018foc    &91bg-like ‘Gold’ &  1.199   &    9.311 $\pm$ 0.47 & 79.993 $\pm$ 3.679 & 17.414 $\pm$ 4.062 & 253.634 $\pm$ 4.041 & 78.078 $\pm$ 2.225 & 232.021 $\pm$ 2.161\\
ZTF18abtnbys &$\cdots$   &91bg-like ‘Gold’ &  0.199   &    9.440 $\pm$ 0.523 & 113.54 $\pm$ 3.726 & 47.809 $\pm$ 3.596 & 233.824 $\pm$ 4.176 & 141.216 $\pm$ 2.191 & 354.292 $\pm$ 2.203\\
ZTF18acrcetn &2018jag    &91bg-like ‘Gold’ &  0.828   &   11.173 $\pm$ 0.347 & 151.745 $\pm$ 8.232 & 76.821 $\pm$ 7.579 &   $\cdots$ & 169.191 $\pm$ 7.408 & 413.613 $\pm$ 29.916\\
ZTF18acybqhe &2018efm    &91bg-like ‘Gold’ &  4.241   &    9.261 $\pm$ 0.372 & 102.062 $\pm$ 2.514 & 45.82 $\pm$ 2.718 & 162.441 $\pm$ 5.224 & 137.024 $\pm$ 1.213 & 221.728 $\pm$ 1.329\\
ZTF18adathpb &2018ldz    &91bg-like ‘Gold’ &  1.814   &    9.04 $\pm$ 1.163 & 82.778 $\pm$ 3.638 & 39.774 $\pm$ 4.085 & 194.826 $\pm$ 3.061 & 101.483 $\pm$ 2.716 & 354.943 $\pm$ 2.908\\
ZTF19aaafica &2019be     &91bg-like ‘Gold’ & -4.975   &   8.705 $\pm$ 0.981 & 71.707 $\pm$ 1.955 & 38.439 $\pm$ 2.001 & 292.39 $\pm$ 2.869 & 87.541 $\pm$ 1.186 &  $\cdots$\\
ZTF19aaaonuk &2019cp     &91bg-like ‘Gold’ &  4.259   &   10.432 $\pm$ 0.811 & 154.165 $\pm$ 2.946 & 51.24 $\pm$ 3.116 & 285.817 $\pm$ 3.57 & 146.902 $\pm$ 1.346 & 355.419 $\pm$ 1.98\\
ZTF19aaejslw &2019afa    &91bg-like ‘Gold’ &  0.249   &   9.988 $\pm$ 0.902 & 118.99 $\pm$ 2.002 & 58.091 $\pm$ 2.162 & 164.258 $\pm$ 4.042 & 108.981 $\pm$ 0.881 & 322.025 $\pm$ 0.991\\
ZTF19aaeopqn &2019abk    &91bg-like ‘Gold’ & -4.782   &   10.617 $\pm$ 0.736 & 122.27 $\pm$ 1.923 & 43.952 $\pm$ 2.148 & 247.085 $\pm$ 3.616 & 191.199 $\pm$ 0.951 & 412.224 $\pm$ 0.86\\
$\vdots$ & $\vdots$    & $\vdots$ &  $\vdots$   &  $\vdots$ & $\vdots$ & $\vdots$ & $\vdots$ & $\vdots$ & $\vdots$ \\
\hline

\end{tabular}
}
\tablefoot{\tiny{(1) It refers to the source, from which the values of the time of maximum light are obtained.}}
\end{table}

\end{appendix}

\end{document}